\documentclass[12pt]{article}
\usepackage{graphicx}% Include figure files
\newcommand{\sss}{\scriptscriptstyle}

\newcommand {\be}{\begin{equation}} % start equation
\newcommand{\ee}{\end{equation}}    % end equation

\def\dds{\frac{\partial}{\partial s}}
\def\dds1{\frac{\partial}{\partial s_1}}

\def\vti{v_{{\sss T}i}}

\def\d{d\kern-0.8 ex\vrule height 1.3 ex depth-1.24 ex width 0.7 ex
\kern 0.15 ex}
\def\D{D\kern-1.7 ex\vrule height .87 ex depth-0.8 ex width 0.7 ex
\kern 0.95 ex}

\begin{document}

\baselineskip 18 pt
\begin{center}
  {\Large{\bf On  the Alfv\'{e}n wave cut-off  in partly ionized collisional plasmas}}
\vspace{0.5cm}

{\bf   J. Vranjes}$^1$ and {\bf M. Kono}$^2$\\
       $^1$Institute of Physics Belgrade, Pregrevica 118, 11080 Zemun, Serbia,\\ Email: jvranjes@yahoo.com\\
 $^2$JFaculty of Policy Studies, Chuo University, Tokyo, Japan,\\ Email: kono@fps.chuo-u.ac.jp

\end{center}
 \thispagestyle{empty}

{\bf   Abstract:}  The cut-off  of  the Alfv\'{e}n  wave,  caused by plasma collisions with    neutrals in multi-component partially ionized plasmas, is discussed.
  Full     multi-component theory is used, and similarities and differences regarding the classic magnetohydrodynamic  theory are presented.
  It is  shown that the cut-off in partially ionized plasma   in principle may remain the same as predicted in  classic magnetohydrodynamic works, although  multi-component theory also yields  some essential differences. Due to electric field, the ion motion is intrinsically two-dimensional and this results in  additional forced oscillations of neutrals. One new small parameter, containing  the ion inertial length, appears in the multi-component theory. This new small parameter is missing in the magnetohydrodynamic description, and it turns out that for some parameters it may be greater than the ions-to-neutrals density ratio which is the only small parameter in the magnetohydrodynamic description. Due to this the  Alfv\'{e}n  wave  behavior can become much different as compared to classic magnetohydrodynamic results.   It is shown also that in plasmas with unmagnetized ions,  Alfv\'{e}n  waves cannot be excited. This by all means applies to the solar      photosphere where the ion collision frequency may be far above the ion gyro-frequency.

\section{\label{s1} Introduction }%First-level heading:\protect\\ The line
%break was forced \lowercase{via} \textbackslash\textbackslash}% ****** Start of file aipsamp.tex ******
%

The magnetohydrodynamic (MHD) description of the cut-off in partially ionized plasma  may be found in the work  \cite{kp} and in many subsequent works \cite{pud, kum, sol}.
In such a MHD description,  the electron and ion components are treated as one single fluid and neutrals as the second one. Friction between the single-fluid plasma on  one side, and neutrals on  the other, causes damping of the  Alfv\'{e}n  wave and this damping increases with the increased number of neutrals. The wave eventually becomes completely non-propagating when the amount of neutrals reaches certain critical value. Alternatively, the same effect appears for increasing the wave-length or wave-period because a plasma particle suffers more and more collisions within one wave oscillation.
 However,  by   increasing the  number of neutrals (or for a greater wave-length or wave-period) the mode is shown to  re-appear again and it may propagate with a very weak damping. In such a regime the plasma and neutrals are collisionally so strongly coupled that there is no friction between them any longer and they move together as a true single fluid.

However, it is frequently overlooked that in this strongly coupled regime, in which neutrals participate in the wave motion, the total fluid density is much greater. Some initial small electromagnetic perturbations  involve the motion of charged species first, and only after some mean collisional time the neutrals are set into motion as well. This all may happen only on the account of the initial energy of the perturbation. Therefore, the initial wave amplitude becomes drastically reduced. In other words, the phase speed of the initial perturbation (for time interval shorter than the collisional time) is necessarily the Alfv\'{e}n speed containing  the plasma density only. After the collisional time, the Alfv\'{e}n speed includes total (plasma plus neutrals) density, and the wave amplitude and flux become  drastically reduced. These features have been overlooked recently in  \cite{ts} resulting in the wave flux which is several orders of magnitude larger from  what may be expected in reality \cite{v1}.

In a recent study \cite{zak}   the authors claimed  that i) the cut-off  obtained in  classic MHD works which they cite was un-physical, ii) it was not an intrinsic property of the wave, and iii) it could  naturally be removed when the Hall and inertia terms are taken into account.
We shall show that  the classic MHD cut-off  cannot be removed by these terms. We shall use full three component equations which naturally incorporate all `additional terms' which they propose as extra physics  within their hybrid MHD-fluid model. Several completely  new features of the Alfv\'{e}n  wave appear in a fully multi-component analysis which we present, and these  can not be predicted within the MHD theory. In addition, we shall show that  Alfv\'{e}n  waves cannot be excited in the lower solar atmosphere due to the fact that ions are un-magnetized. This has profound implications on the coronal heating model based on the Alfv\'{e}n  waves that are assumed massively produced by convective motions in the lower solar atmosphere.

\section{\label{b} The   Alfv\'{e}n  wave in partially ionized plasmas}

The physics of partially ionized plasma is considerably richer as compared to the fully ionized one. Some different type of collisions appear, the system becomes intrinsically multi-component, particles may be  lost and re-created, etc. Therefore before discussing the   collisional damping of the Alfv\'{e}n  wave in such an environment it is necessary to provide accurate  collisional cross sections for collisions of interest here. We shall take ions to be protons, and neutrals are parental hydrogen atoms.
\begin{figure}
 \centering
\includegraphics[height=6cm,bb=17 16 273 230,clip=]{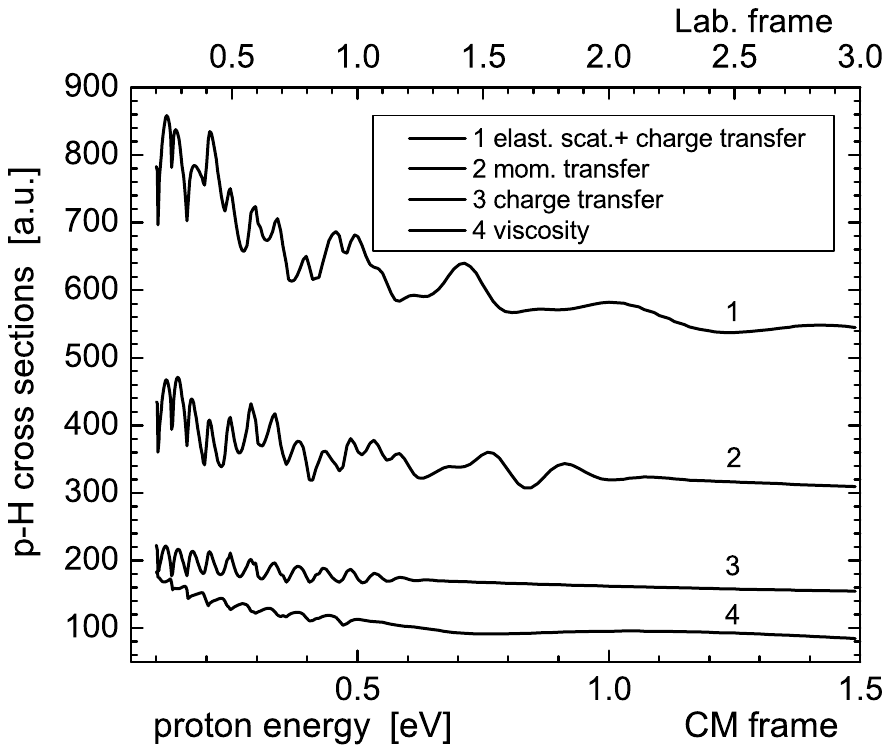}
\caption{\label{f1}Integral cross sections $\sigma_{p{\sss H}}$  (in units a.u.$= 2.8\cdot 10^{-21}$ m$^{2}$)  for proton collisions with neutral hydrogen $H$ in terms of proton energy  and for the model of quantum-mechanically indistinguishable particles, from  \cite{kr1, kr2, kr3, vaa}.  }
\end{figure}

In Fig.~\ref{f1} we give the  cross sections for proton collisions with neutral hydrogen obtained from quantum theory, in the range $0.05-1.5$ eV proton energy in the center of mass (CM) frame (bottom $x$-axis). In the laboratory (plasma) frame the energy range $0.1-3$ eV is given by the top $x$-axis using the transformation formula $E_{lab}= E_{\sss {CM}} (m_1 + m_2)/m_2$, where $m_2$ is the target particle.   %The graph is  from works of \citet{kr1, kr2, kr3}.
 It  provides  the following  important features: a) velocity (energy) dependent cross sections,  b) different cross sections for different phenomena (elastic scattering, momentum transfer, charge exchange, viscosity), c) the charge exchange as a specific sort  of inelastic collisions that cannot be omitted in systems containing both ions and their parental atoms. These  three features   are rarely seen studied in  space plasmas, separately or together. In the present context they have never been studied in the past. It will be shown below that this is the only proper approach.

  In application to the solar atmosphere, the energy dependence  [the feature  a)]  in the present study is equivalent to the altitude (i.e., temperature) dependence in the stratified solar plasma.
 The four different lines [the feature  b)] describe different cross sections:  the sum of elastic scattering  and charge exchange (line 1),  momentum transfer (line 2), charge exchange alone (line 3), and viscosity (line 4). The values presented here are the most accurate that exist. The physics of  this difference between various cross sections  is described partly in a recent work \cite{vaa} and in much more detail in \cite{kr3,glas},   and in \cite{schul}  The quantum-mechanical indistinguishability of  particles used here implies overlapping of particle quantum wave functions at low energies. As described in \cite{kr3}, when the nuclei are identical and the collision energy is relatively low, in the process of collisions it is not possible  to
distinguish the ion which is elastically scattered  from the  ion which originates  from charge transfer unless  they are additionally labeled (e.g., by  their spin). For this reason the line 1 in Fig.~\ref{f1} contains the sum of  the  elastic scattering  and the charge exchange cross sections. This total cross section should be used in the estimate of magnetization of plasma particles, while for example the cross section for momentum transfer is to be used in the friction force terms.  More details on this issue may be found in \cite{vaa}. 

The lines for the momentum transfer  and viscosity cross sections   are obtained after an integration over the scattering angle $\theta$ similar to the line 1, but weighted by $1-\cos \theta$ and $\sin^2 \theta$, respectively. This describes essentially different physics involved in the momentum transfer and viscosity, therefore the different lines in Fig.~\ref{f1}. In the case of viscosity this emphasizes the scattering at the angle $\pi/2$ and de-emphasizes the forward and backward  ones, while the factor $1-\cos \theta$ in the momentum transfer emphasizes the backward scattering angles. Note also that elastic scattering is typically  forward, and charge exchange backward scattering process.      We stress that both the momentum transfer  and the viscosity lines also contain the contribution from the charge exchange effect.  In practical application further in the text, this also implies that in the given approach in describing plasma wave dynamics we shall have  one single  `friction force' term which  determines the total wave damping on neutrals, instead of two separate  terms as typically seen in the  literature, one for friction force caused by elastic collisions and one for the momentum lost (or gained) due to charge exchange.
 This all not only yields much more accurate  results (in view of the most reliable  cross sections we use)  but also considerably simplifies  derivations as will become clear below.

We note that, following the recipe from \cite{kr1}, the commonly used  (classic) elastic scattering differential cross
sections may be obtained as the absolute value of the difference of the total (elastic plus  charge transfer) differential cross section which is in the basis of the model we use here, and charge transfer differential cross section. With this we   can then calculate the integral quantities (i.e.,  momentum transfer and viscosity cross sections) in order  to obtain results for  the right classical limit. As described in \cite{kr1}, this limit implies the model of distinguishable particles.

It should be stressed again that partially ionized plasmas like the lower solar atmosphere contain ions and corresponding parental neutral atoms, so the charge exchange cannot possibly be omitted [the feature c)]. In laboratory conditions in inert gases (helium, neon, argon) the charge exchange cross section is  the largest one \cite{reiz}. In the case of hydrogen it is below the cross section for elastic scattering. However, because of the described specific angle dependence of the charge transfer from one side, and rather different  angle dependence of the other three cross sections  in Fig.~\ref{f1} from the other side, it contributes considerably (and differently) to all three of them. Therefore  the procedure presented here has no alternative, it has been tested in numerous works in the past like in the references  mentioned above, and it gives values in complete agreement with laboratory measurements, e.g., in \cite{varg}.

In what follows we shall also need the momentum  transfer cross section for electron-hydrogen collisions. Some values of the cross section in the energy (temperature) range of interest here  are given in Table~\ref{teh}, from \cite{dal}.

\begin{table}
 \caption{Cross section for momentum transfer for e-H collisions, following \cite{dal}. }\label{teh}
  \centering
  \begin{tabular}{lccccc}
    \hline
    el. energy [eV] &0.2 & $0.3$ & $0.5$ & $0.7$ & $1$ \\
     $\sigma_{eH,mt}$ [$\times 10^{-19}$ m$^2$]&3.7  & $3.1$ & $3$ & $2.8$ & $2.5$\\ 
    \hline
  \end{tabular}
 \end{table}

To proceed with the   Alfv\'{e}n  wave cut-off, we use fully three component theory without any assumptions  following the procedure that may be found in standard textbooks (e.g.\ \cite{ch}) for  the shear Alfv\'en wave.
 We take  $\vec B_0= B_0\vec e_z$, and in such a geometry
both ion and electron fluids oscillate together in the direction of the perturbed magnetic field vector $\vec B_1=B_1 \vec
e_y$. This is due to the $\vec E_1 \times \vec B_0$ drift, which separates neither charges nor masses. The direction of the electric field is determined by the Faraday law. The wave is  further sustained by the additional polarization drift $\vec v_{pj}= (m_j/q_jB_0^2) \partial \vec E_1/\partial t$ and the consequent Lorentz force $j_x
\vec e_x \times \vec B_0$, which is again in the $y$-direction and has a proper phase shift. The
polarization drift appears as a higher order term due to $|\partial/\partial t|\ll \Omega_i$. It introduces the ion
inertia effects and if it is neglected, then the Alfv\'en wave vanishes. The wave thus develops at time scales far greater than the ion gyro-rotation time, and equivalently at spatial scales far exceeding the ion gyro-radius.   The mode is fully described by the  wave equation which is obtained by combining Amp\`{e}re and  Faraday law equations
\be
\nabla\times \left(\nabla\times  \vec E\right) + \mu_0 \frac{\partial \vec j}{\partial t}=0, \label{bwe} \ee
where the displacement current is omitted as appropriate for phase speed  far below the speed of light. The plasma current  is calculated by using the linearized momentum equations for ions and electrons
\[
m_in_0\frac{\partial \vec v_i}{\partial t}  = e n_0\left(\vec E + \vec v_i\times \vec B_0\right) - m_i n_0 \nu_{ie} (\vec v_i - \vec v_e)
\]
\be
 - m_i n_0 \nu_{in} (\vec v_i - \vec v_n),\label{be1a}
 \ee
\[
m_en_0 \frac{\partial \vec v_e}{\partial t}  = -e n_0\left(\vec E + \vec v_e\times \vec B_0\right)- m_e n_0 \nu_{en} (\vec v_e - \vec v_n)
\]
\be
 - m_e n_0 \nu_{ei} (\vec v_e - \vec
v_i),\label{be1b}
 \ee
and the corresponding equation for neutrals
 \be
 \frac{\partial \vec v_n}{\partial t}  =  -\nu_{ne}(\vec v_n-\vec
v_e) -\nu_{ni}(\vec v_n-\vec v_i).
 \label{be1c}
\ee
Here, the static quasi-neutral equilibrium without macroscopic flows is assumed and the equilibrium quantities are denoted by the subscript $0$.
The last terms in Eqs.~(\ref{be1a}, \ref{be1c}) describe the momentum change due to $p-H$  collisions and it includes the contribution from both
elastic collisions and charge exchange, i.e., the corresponding collision frequency includes the momentum transfer collision cross section $\sigma_{mt}$ (line 2 from Fig.~\ref{f1}), $\nu_{in}=\sigma_{mt} n_{n0} \vti$. The momentum conservation in Eq.~(\ref{be1c}) implies
\be
\nu_{ni}=m_i n_{i0} \nu_{in}/(m_n n_{n0}), \quad \nu_{ne}=m_e n_{e0} \nu_{en}/(m_n n_{n0}). \label{mc}
\ee
 Further in the text we shall use $m_n=m_i$. In  charge exchange collisions the number of protons and neutrals does not change so there is no source-sink term in the continuity equation and the latter in the cold plasma case becomes redundant. However, a source-sink momentum change appears  in the momentum equation through $\sigma_{mt}$ which contains contribution from the charge transfer. In such collisions  a proton takes over one electron from the hydrogen atom. The latter then becomes charged and is consequently directly involved in the wave motion, yet it does not share the same momentum as the other protons and some wave energy must be spent in order to set it into motion. Similarly, the described proton which becomes neutral atom takes away a part of the momentum previously gained from the wave. All these effects are now self-consistently introduced.  It should be stressed that very frequently  in the literature the friction force term appears separately from the term which describes the momentum change by the charge exchange, see for example in works  \cite{cat} and \cite{hel}. The two forces are given by $\vec F_f=m_i n_i \nu_1 \vti (\vec v_i-\vec v_n)$ and  $\vec F_{cx}=m_i n_n\nu_2\vti (\vec v_i-\vec v_n)$ where $\nu_1=\sigma_1 n_n\vti$, $\nu_2=\sigma_2 n_i\vti$, and   $\sigma_{1, 2}$ should be calculated following the model of distinguishable particles as described  previously in this section.  The total force reduces to $\vec F=m_i n_i (\sigma_1+ \sigma_2) \vti (\vec v_i-\vec v_n)$  where now the sum of the cross sections corresponds to our $\sigma_{mt}$.  In  such an approach  equations are  more complicated and  a great care is needed so that  the corresponding collision frequencies  are calculated correctly.

Due to the absence of the source-sink terms in the continuity equations, and also in order to keep the present model close to those  from \cite{kp} and \cite{zak}, the density perturbations, thermal effects and viscosity are neglected. So the number densities in Eqs.~(\ref{be1a}-\ref{be1c}) describe the equilibrium quantities and we shall take care that  we are in the correct wave phase speed range  to satisfy such an assumption (roughly speaking in the small plasma-$\beta$ limit). Hence, Eqs.~(\ref{bwe}-\ref{be1c}) together with the quasi-neutrality  condition which is used $n_i=n_e=n_0$ make a  closed set.

We stress that a complete comparison of our model and results with those available in the literature (including the above cited two references) is not possible due to the following reasons. Even if we would disregard the charge exchange, there remains the fact that typical  models,  presently (and previously) widely used   in the literature, make no distinction between the cross sections for momentum transfer and elastic scattering. However, these two cross sections are different, and the charge exchange additionally (and differently) contributes to both of them. These facts make the theory we provide here much more accurate in comparison  to what may be found in the literature.

%\subsection{ MHD derivation of Kulsrud and Pearce.}\label{kp}

\subsection{\label{kp}Classic MHD derivation}

\begin{figure}%[!htb]
   \centering
  \includegraphics[height=6cm,bb=16 14 262 216,clip=]{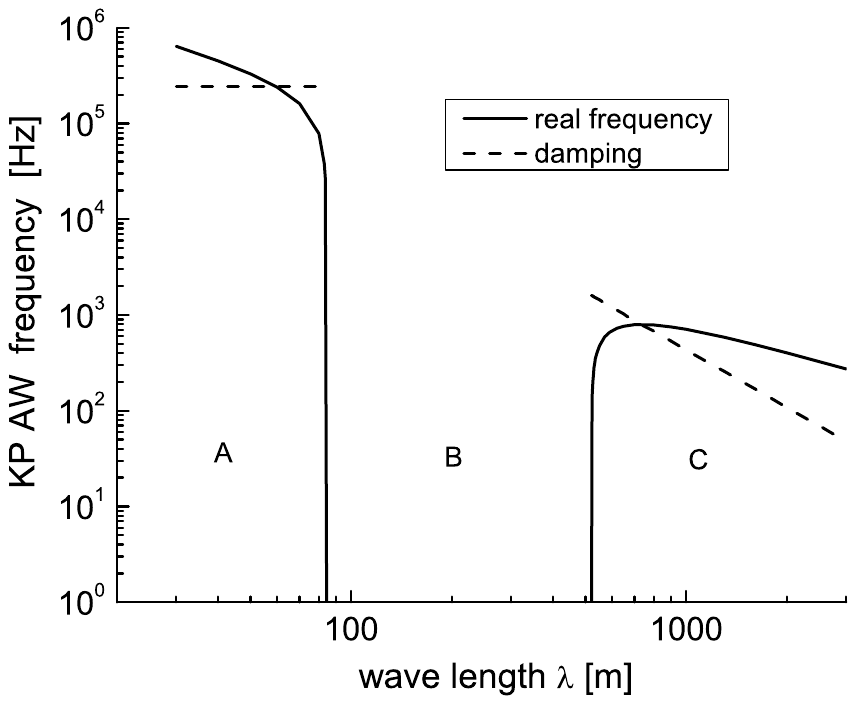}
      \caption{ \label{kp-new}Kulsrud-Pearce solution $\omega=\omega_r- i \gamma$ of Eq.~(\ref{kp3})   for parameters given in the text.
      }
       \end{figure}

In order to obtain the classic result of \cite{kp} and for comparison with our results later in text, the electron and
ion momentum equations are summed, their mutual friction vanishes, there appears the current in the Lorentz force term
which is then expressed through the magnetic field from the Amp\`{e}re law. The resulting momentum equation is
\[
m_i n_i \frac{\partial \vec v_i}{\partial t} + \underline{m_e n_e \frac{\partial \vec v_e}{\partial t}} = \underline{e(n_i-n_e)\vec E} + \frac{1}{\mu_0} \left(\nabla\times \vec B\right)\times \vec B
\]
\be
- m_i n_i \nu_{in} \left(\vec v_i-\vec v_n\right) -  \underline{ m_e n_e \nu_{en}\left(\vec v_e-\vec v_n\right)}. \label{kp1}
\ee
The underlined terms are further omitted, and what remains  is combined with the neutral equation (\ref{be1c})
 where the electron contribution is omitted as well. The system is closed with  the induction equation
\be
  \frac{\partial \vec B}{\partial t}=\nabla \times (\vec v\times \vec B), \label{kp2}
  \ee
  where for $\vec v$ the ion speed is used. This yields the well known cubic dispersion equation \cite{kp}:
  \be
  \omega^3 - \omega k^2 c_a^2 + i\nu_{in} \left[\left(1+ \frac{n_0}{n_{n0}}\right) \omega^2 - k^2 c_a^2 \frac{n_0}{n_{n0}}\right] =0.\label{kp3}
  \ee
  Dispersion equation (\ref{kp3}) contains one small parameter $n_0/n_{n0}$  (here and further in the text the medium is assumed as weakly ionized). It is solved numerically for the following arbitrary parameters: $B_0=0.05$ T,
   $n_0=1.1 \cdot 10^{17}$ m$^{-3}$, $n_{n0}=6.8 \cdot 10^{19}$ m$^{-3}$, $T=5755$ K.  This yields
   $\Omega_i=4.8\cdot 10^6$ Hz, and with the help of Fig.~\ref{f1} we have $\nu_{in}=4.88\cdot 10^5$ Hz
    for the momentum transfer.  Note that these parameters in fact describe plasma in the chromosphere at
     altitude 900 km, although  the assumed magnetic field may be too strong. However,  this is immaterial
      for the present purpose as we only  want to describe basic features of the classic MHD result, and
       to compare it with the multi-component description later in the text.    The Kulsrud-Pearce (KP)  complex frequency
        is calculated from Eq.~(\ref{kp3}) and presented in Fig.~\ref{kp-new}. The  real part of the frequency  vanishes
        above the critical wave length  $\lambda_{c1}\simeq 84.35$ m, and re-appears again for wave lengths above
        $\lambda_{c2}\simeq 521.5$ m.  The damping in the short wave length region $A$ is almost constant
        $\gamma\simeq \nu_{in}/2 \simeq 2.44\cdot 10^5$ Hz as can be deduced also analytically from Eq.~(\ref{kp3}).
        In the upper propagation window $C$
  in most of the spectrum we have roughly $\gamma\sim 1/\nu_{in}$, and this too can be analytically deduced from Eq.~(\ref{kp3}).

  Several comments are in order. With the assumed geometry $\vec B_1\equiv B_y\vec e_y$, it is seen that after neglecting
  the underlined terms in Eq.~(\ref{kp1}), the motion of the ion  center of mass is strictly one-dimensional
   because $\left(\nabla\times \vec B_1\right)\times \vec B_0$ is parallel to $\vec v_{i1}$. This is completely different as compared with
  our multi-component model later in the text where the motion of the ion center mass is essentially two-dimensional.
  The reason for difference is clearly the fact that the electric field vanishes from Eq.~(\ref{kp1}).  `Vanishing' of
  electric field only tells us that {\em electric force}  (as a vector), when acting on total plasma (e+i) is zero. Physically
  this means simply that opposite charges suffer equal force with opposite direction and nothing else, while in fact the electric
   force acting on ions and electrons  separately is still present. Observe that in  Eq.~(\ref{kp1}) the contribution of the electric
   force vanishes only because of quasi-neutrality and not because of actual absence of the electric field. In other words, the
   electric field is always present and the motion of electrons and ions  is fully two-dimensional in perpendicular direction, yet
   this information is lost in the KP-MHD equation. The presence of the electric field is logical because it is there  due to electromagnetic
   perturbation, and this even without plasma. In other words, magnetic field is disturbed in $y$-direction  and the Faraday law then
    tells us  that there is electric field in $x$-direction (even without plasma).

  Observe also that the KP model is based on an ideal frozen-in environment for the magnetic field [this is incorporated
   in the induction equation (\ref{kp2})]. This also implies that  the only current which was  initially in the Lorentz
    force $\sim \vec j\times \vec B$, was in fact the ion polarization drift current $\vec v_{p,i}\sim \vec E_x\vec e_x$ [this is
     because the other, more dominant, $\vec E\times \vec B$-drift  is the same for electrons and ions, so the plasma motion in
      $y$-direction is current-less]. Hence, although the starting momentum equation {\em implies} two-dimensional motion (that
      is the polarization drift in $x$-direction and  $\vec E\times \vec B$-drift in $y$-direction), from   Eq.~(\ref{kp1}), without
      underlined terms,  we have a purely one-dimensional ion dynamics. This is just one out  of many  contradictions that are inherent to the MHD model, some others may be seen  in the work  \cite{bel}.

       %One obvious way to reintroduce two-dimensionality  of ions and electrons  is to abandon the ideal frozen-in equation
       % (\ref{kp2}), which is clearly far from reality in a heavily collisional environment \citep{kris, pan2, pan}, and to use full response of electrons and
        % ions. Note that keeping the Hall term $\vec j\times \vec B_0$ does not solve the problem because this  is again a vector in
        %  $y$-direction and the motion remains one-dimensional.

\subsection{\label{ap}Approximate analytical  approach}

Using an approximate method we shall now demonstrate that there can be no Alfv\'{e}n  wave
 if ions are un-magnetized. Although self-evident, this fact is  frequently ignored  and Alfv\'{e}n
  waves have been studied in lower solar atmosphere where ion collision frequency (above $10^9$ Hz) far
  exceeds the ion gyro-frequency, and  where these waves cannot possibly be excited.

Because of lengthy expressions, in this section  we may take ion collisions with neutrals most dominant,
 which is easily satisfied in the lower solar atmosphere  \cite{vaa}. From the ion equation (\ref{be1a})
  we may first express the perpendicular ion speed from the friction force term:
 \be
 \vec v_{\bot i}=\frac{e \vec E_1}{m_i \nu_{in}} + \frac{e }{m_i \nu_{in}} \vec v_{\bot i}
 \times \vec B_0 -\frac{1}{\nu_{in}} \frac{\partial \vec v_{\bot i}}{\partial t} + \vec v_{\bot n}. \label{vi1}
 \ee
 Next, we  make vector product $\vec e_z\times$ of Eq.~(\ref{be1a}) and express the ion perpendicular
 speed from the Lorentz force term:
 \[
 \vec v_{\bot i}=-\frac{1}{B_0} \vec e_z \times   \vec E_1 + \frac{\nu_{in} }{\Omega_i} \vec e_z\!\times\!
 \vec v_{\bot i} + \frac{1}{\Omega_i} \frac{\partial}{\partial t} \vec e_z \times \vec v_{\bot i}
 \]
 \be
 - \frac{\nu_{in}}{\Omega_i}\vec e_z\times \vec v_{\bot n}. \label{vi2}
 \ee
The neutral speed is eliminated by using Eq.~(\ref{be1c}) which yields
 \be
 \vec v_{\bot n}=\frac{i \nu_{ni}}{\omega+ i \nu_{ni}} \vec v_{\bot i}. \label{neutp}
 \ee
 Here, the momentum transfer due to neutral-electron collisions is omitted as higher order in comparison to
 their collision with ions. The  two expressions for the ion speed are made equal,  and the  vector product
  $\vec e_z\times$ is again applied on the resulting equation. This yields the following recurrent formula
  with the small parameter $|\partial/\partial t|/\Omega_i\ll 1$:
\[
\vec v_{\bot i} = \alpha_i \left( -\frac{1}{B_0} \vec e_z\times  \vec E_1 + \frac{\nu_{in}}{\Omega_i} \frac{\vec E_1}{B_0} +
\frac{1}{\Omega_i} \vec e_z\times \frac{\partial \vec v_{\bot i}}{\partial t} \right.
\]
\be
\left.- \frac{\nu_{in}}{\Omega_i^2} \frac{\partial \vec v_{\bot i}}{\partial t} +
\frac{i \nu_{in}^2}{\Omega_i^2}\frac{\nu_{ni}}{\omega_n} \vec v_{\bot i} -
\frac{i \nu_{in} \nu_{ni}}{\Omega_i \omega_n} \vec e_z \times \vec v_{\bot i} \right).\label{vi3}
\ee
Here, $\omega_n=\omega+ i \nu_{ni}$, and $\alpha_i=1/(1+ \nu_{in}^2/\Omega_i^2)\leq 1$. Eq.~(\ref{vi3})
will further be discussed  for  two separate cases.

\subsubsection{\label{smal}Small ratio $\nu_{ni}/|\omega_n|$. Explicit absence of Alfv\'{e}n  wave for un-magnetized ions}

In the case when  $\nu_{in}/\Omega_i$ is arbitrary but finite, the first two terms on the right-hand side in Eq.~(\ref{vi3}) may be assumed as  leading order,
 and applying standard approximate   procedure they can be used to replace $\vec v_{\bot i}$  in the remaining four terms.
In the same time  the  following  small ratio is assumed     in the last two terms in Eq.~(\ref{vi3}):
 \be
 \nu_{ni}/|\omega_n|\ll 1. \label{frat}
 \ee
   This condition can  easily be satisfied in view of Eq.~(\ref{mc}) and for a weakly ionized environment where $n_{0}/n_{n0}\ll 1$, although it is not so general. With this the recurrent formula
   (\ref{vi3}) remains  valid and after  a few steps this yields
  \[
\vec v_{\bot i} = \alpha_i \left\{ -\frac{ \vec e_z\times  \vec E_1}{B_0}\left( 1+ \frac{i 2 \alpha_i \nu_{in} \omega}{\Omega_i^2} +
\frac{i 2 \alpha_i \nu_{in}^2\nu_{ni}}{\Omega_i^2 \omega_n} \right) \right.
\]
\be
\left.
+\frac{ \vec E_1}{B_0}\! \left[\! \frac{\nu_{in}}{\Omega_i} - \frac{i \alpha_i \omega}{\Omega_i}\! \left(\!1- \frac{\nu_{in}^2}{\Omega_i^2}\!\right)\!
- \frac{i \alpha_i \nu_{in} \nu_{ni}}{\Omega_i \omega_n} \! \left(\!1- \frac{\nu_{in}^2}{\Omega_i^2}\!\right)\!\right]\!\right\}.
\label{vi4}
\ee
For the present  purpose it may be good enough to omit electrons and their collisions completely. For large time and space scales of interest here, the  mobile electrons will closely follow the ion dynamics. On the other hand, their contribution to dragging (by friction) of the heavy background of neutrals is small in any case, as already assumed in Eq.~(\ref{neutp}). This will be confirmed later in the text, see Fig.~\ref{4-omega}. With such an approach  we shall remain  as close to the classic MHD theory as possible.

To make this point more clear,  derivations for electrons can be repeated in a similar manner  with the  distinction  that  the left hand side of Eq.~(\ref{be1b}) is omitted, which is fully justified in view of the mass difference.
This yields
\[
\vec v_{\bot e}=\alpha_e \left(-\frac{ \vec e_z\times  \vec E_1}{B_0} - \frac{\nu_e}{\Omega_e} \frac{ \vec E_1}{B_0} + \frac{\nu_{en}}{\Omega_e} \vec e_z\times \vec v_n + \frac{\nu_e\nu_{en}}{\Omega_e^2} \vec v_n\right.
\]
\be
\left.
 \frac{\nu_{ei}}{\Omega_e} \vec e_z\times \vec v_i + \frac{\nu_e\nu_{ei}}{\Omega_e^2} \vec v_i\right). \label{dod1}
\ee
Here, $\alpha_e=1/(1+ \nu_{en}^2/\Omega_e^2)\simeq 1$, $\nu_e=\nu_{ei}+ \nu_{en}$. In Eq.~(\ref{dod1})  collisions with both ions and neutrals are formally kept, and in principle the neutral and ion speeds here  should be calculated from full Eqs.~(\ref{be1a},\ref{be1c}) to satisfy conservation laws, but this will not be necessary as shown  below.

For the wave equation (\ref{bwe}) we need only $x$-component of Eqs.~(\ref{vi4}, \ref{dod1}). We may also take $\nu_{en}> \nu_{ei}$, this will not affect  generality of conclusions below.  So now we may  compare  the leading $x$-term from the electron equation   (that is the second term) with any of the $x$-terms from the ion  equation (and we take the second term again). These two   yield the following contribution to the current in the wave equation
\[
 \left(\alpha_i\frac{\nu_{in} }{\Omega_i} +\alpha_e \frac{\nu_{en}}{\Omega_e}\right)\frac{E_x}{B_0}= n_{n0} (\alpha_i \sigma_{in}  \rho_i + \alpha_e\sigma_{en} \rho_e) \frac{E_x}{B_0}.
\]
 From Table~\ref{teh} and from Fig.~\ref{f1} (see the line 2 there), for energies of interest here we have always $\sigma_{in}> \sigma_{en}$, while in the same time  $\rho_i\gg \rho_e$. Hence, as long as $\alpha_i>\alpha_e(\rho_e/\rho_i) (\sigma_{en}/\sigma_{in})$ the electron contribution can completely be  neglected. This condition is easily satisfied if we allow  the ratio  $\nu_{in}^2/\Omega_i^2$ to be finite (with any of the two possibilities: $\nu_{in}^2/\Omega_i^2> 1, < 1$), so that  $\alpha_i\leq 1$, while  $(\rho_e/\rho_i) (\sigma_{en}/\sigma_{in})\ll 1$ and  $\alpha_e\leq 1$ or  $\alpha_e \simeq  1$.

Observe that keeping electron inertia term  would yield electron polarization current in the $x$-direction, but it is  proportional to the electron mass and  indeed negligible as we assumed above [see also Eq.~(\ref{epol}) in the following  section].

Consequently, we may indeed  proceed by omitting electrons and for the ion $x$-component we have
\be
v_{ix} = \frac{\alpha_i \vec E_1}{B_0}   \left[ \frac{\nu_{in}}{\Omega_i} - \frac{i\alpha_i}{\Omega_i}  \left(1- \frac{\nu_{in}^2}{\Omega_i^2}\right)\left(\omega + \frac{\nu_{in}\nu_{ni}}{\omega_n}\right)\right].
\label{vi5}
\ee
From Eqs.~(\ref{bwe}, \ref{vi5})
the following approximate dispersion equation is obtained:
\[
\omega^2 \left(\nu_{in}^2-\Omega_i^2\right) - i \omega \nu_{in} \left(\nu_{in}^2+\Omega_i^2\right)
+ k^2 \lambda_i^2 \left(\nu_{in}^2+\Omega_i^2\right)^2
\]
\be
+ \nu_{in}\nu_{ni} \left(\nu_{in}^2-\Omega_i^2\right)=0.
 \label{cc1}
\ee
 In the absence of collisions this yields  a real Alfv\'{e}n mode $\omega^2= k^2 c_a^2$.
On the other hand, in particular case when $\nu_{in}=\Omega_i$ we obtain
\[
\omega=- i 2 k^2\lambda_i^2\nu_{in}, \quad \lambda_i=c/\omega_{pi}.
\]
Hence, no real mode exists in this case.

 In the presence of collisions in general, Eq.~(\ref{cc1}) has no real solutions if
ions are   {\em un-magnetized}:
 \be
 \nu_{in}\ge\Omega_i. \label{st}
 \ee
 On the other hand, for weakly magnetized ions  $\nu_{in}<\Omega_i$ from Eq.~(\ref{cc1}) we obtain very approximately
 that real solutions are possible provided that
 \be
\frac{\Omega_i}{\nu_{in}} > \frac{1}{2k\lambda_i}. \label{cc22}
\ee
The condition  (\ref{cc22}) determines the Alfv\'{e}n wave cut-off for waves satisfying the condition (\ref{frat}).

 The condition (\ref{frat}) clearly includes the case $\nu_{ni}=0$  as well. From our  starting equation (\ref{be1c}) with omitted electron effects   we see that  this is equivalent to assuming neutrals as a static background (fairly well justified for short wavelengths). In this case  Eq.~(\ref{cc1}) reveals that there can be no Alfv\'{e}n wave  for  unmagnetized ions $\nu_{ni}>\Omega_i$.

\subsubsection{\label{arb}Arbitrary ratio $\nu_{ni}/|\omega_n|$}

We shall now repeat the approximate procedure  with the only condition
\be
\frac{|\omega|}{\Omega_i}\ll 1.
\label{con2}
\ee
 In Eq.~(\ref{vi4}) we eliminate the terms with the vector product $\vec e_z\times
\vec v_{\bot i}$ by using Eq.~(\ref{vi2}). After a few steps this yields
\[
\vec v_{\bot i} =- \frac{\delta_i}{B_0} \vec e_z\times \vec E_1 + \delta_i \frac{\omega}{\Omega_i} \left(\frac{\nu_{in}}{\omega_n} - i\right)\frac{\vec E_1}{B_0}
\]
\be
 +
2 \delta_i  \frac{\omega}{\Omega_i} \frac{\nu_{in}}{\Omega_i}\left(\frac{\nu_{ni}}{\omega_n} + i\right) \vec v_{\bot i}, \quad \delta_i=\frac{\alpha_i}{\beta_i \gamma_i},  \label{cc2}
\ee
\[
\beta_i=1- i \alpha_i \frac{\nu_{ni}}{\omega_n} \frac{\nu_{in}^2}{\Omega_i^2}, \quad \gamma_i=1-\frac{\alpha_i}{\beta_i}\frac{\nu_{ni}}{\omega_n} \frac{\nu_{in}^2}{\Omega_i^2} \left(i+ \frac{\nu_{ni}}{\omega_n}\right).
\]
Because of the small ratio   (\ref{con2}), the first term on the right-hand side in the recurrent formula (\ref{cc2}) is the leading order one, so it is used in the last term. From the resulting equation we need only $v_x$  in  the wave equation
\[
k^2 E_x- i e n_0 \mu_0 \omega v_x=0.
\]
Hence,
\[
v_x=    \delta_i \frac{\omega}{\Omega_i} \left(\frac{\nu_{in}}{\omega_n} - i\right)\frac{E_1}{B_0}
\]
is used in the wave equation yielding  the following approximate dispersion equation
\[
\omega^4+ i \omega^3 \left(\nu_{in} + 2 \nu_{ni}\right) - \omega^2 \left(\nu_{in} \nu_{ni}+ \nu_{ni}^2 + k^2 \lambda_i^2 \nu_{in}^2 + k^2 c_a^2\right)
\]
\be
- 2 i \nu_{ni} k^2 c_a^2 \omega + k^2 c_a^2 \nu_{ni}^2=0.\label{cc3}
\ee
 Contrary to \cite{kp} result (\ref{kp3}), the obtained equation is 4th order.
  It describes a) the Alfv\'{e}n wave, and b) some low frequency forced neutrals' (FN) oscillations due
  to their coupling with plasma.   These FN oscillations appear only due to the fact that the induced neutral
   motion is two-dimensional.    Setting $v_{ny}=0$ or $v_{nx}=0$   yields a third order equation instead
   of Eq.~(\ref{cc3}), which is then equivalent to the result of \cite{kp} and \cite{pud}, so the
    FN collisional mode vanishes.  In similar studies with friction related to the ion acoustic (IA)
    waves \cite{v10} this extra collisional mode does not appear; the neutral response for longitudinal
    IA waves is one-dimensional.   We stress that FN mode describes  forced
    oscillations and   not  a normal mode in a neutral gas,  it is  caused by coupling of neutrals with plasma but propagates
     independently of plasma modes and this only in the neutral gas (see also \cite{gg}). To get a rough glimpse of the FN mode we may write Eq.~(\ref{cc3}) as
     $D(\omega_r+ i \gamma, k)\equiv D_r+ i D_{im}=0$. Knowing that the spectrum (number  of modes) must be determined by the real part,
       we may set $D_r=0$ which then yields an obvious Alfv\'{e}n wave part and additional terms describing the FN mode, and some coupling terms.
       Assuming that the AW part is weakly affected by the FN mode and that frequencies
     of the two modes are well separated,  in the equation $D_r=0$ we may
     set the Alfv\'{e}n part separately  equal to zero. The remaining terms yield  very roughly
     \be
     \omega^2\simeq \frac{\epsilon^2 k^2 c_a^2 }{\epsilon + k^2\lambda_i^2}, \quad \epsilon=\frac{n_0}{n_{n0}}. \label{fnn}
     \ee
     This expression describes the FN mode accurately only in the long wavelength limit and where the AW is absent, see Sec.~\ref{dkp}.
     In the short wavelength regime it gives the
     frequency which is far from actual  values because the mode is very strongly damped and many essential terms from the imaginary part of (\ref{cc3})
     are missing in the given expression. But note the presence of two small terms in Eq.~(\ref{fnn}), the previously obtained (within the MHD approach) $\epsilon$, and the new one $k^2\lambda_i^2$. Further in the text it will be shown that the interplay of these two is crucial for the Alfv\'{e}n wave behavior.

     Observe that the BGK integral used in derivations is the same
      for both Kulsrud-Pearce and our multi-component model, therefore  it should give similar results in
      the two descriptions. However, it will be shown in Sec.~\ref{dkp} that this is dependent on parameters and in some cases the results are
      essentially different.

 Eq.~(\ref{cc3}) will be discussed together with Eqs.~(\ref{ni}, \ref{de}) which are presented  later in the text.
 It will be shown that, depending on parameters,   the two propagation windows  A and C from Fig.~\ref{kp-new}  and the AW cut-off may remain intact as predicted in \cite{kp},   contrary to recent  claims \cite{zak}.

\subsection{\label{in}Ion-neutral plasma without approximations}

In the wave equation we need $x$-component of the perturbed speed so neglecting contribution of electrons in
starting equations  Eqs.~(\ref{bwe}-\ref{be1c})  and keeping all remaining terms, thus  without approximations
based on the recurrent formula (\ref{vi3}), yields the following dispersion equation:
\[
\omega^4 \left(1+ k^2 \lambda_i^2\right) + i \omega^3 \left[ \nu_{in}+ 2 \nu_{ni} + 2 k^2 \lambda_i^2 \left(\nu_{in}+  \nu_{ni}\right)\right]
\]
\[
- \omega^2\!\left[\nu_{in} \nu_{ni} + \nu_{ni}^2+ k^2 c_a^2 + k^2 \lambda_i^2\! \left(\nu_{in}+  \nu_{ni}\!\right)^2\right]\! - i 2 k^2 c_a^2 \nu_{ni} \omega
\]
\be
 + \, k^2 c_a^2 \nu_{ni}^2=0. \label{ni}
\ee
Clearly this is very similar to Eq.~(\ref{cc3}) obtained above from the recurrent formula, with a few additional but unessential
 terms, and it  describes  again the  Alfv\'{e}n wave modified by collisions,  and the FN  mode. In view of omitted electrons, this is also equivalent to
the KP equation (\ref{kp3}), but compare  the order of these two equations.

We stress again that although both electrons and ions  contribute to these neutral oscillations, the electron contribution is
completely negligible as will be shown quantitatively in the following section. Eq.~(\ref{ni}) is solved numerically in the following section.

\subsection{\label{full}Full multi-component model with complete electron contributions}

We now perform derivations  for full three-component case with  electron dynamics and all collisions included.
Note  that   following the usual procedure, Eqs.~(\ref{bwe}-\ref{be1c}) can be transformed and combined yielding the standard MHD equations used in
\cite{kp}, and in \cite{zak}, including the generalized Ohm's law with additional terms which, according
to \cite{zak} remove the cut-off. Hence, all these `additional' terms  are naturally already present in our fully multi-component set of equations (\ref{bwe}-\ref{be1c}). Discussion about  these additional terms in fact  serves the purpose only within the  MHD modeling which works well within  known limits, but as an approximate theory it omits some physics,  and such additions are used to re-introduce back the physics which is removed initially by reducing natural multi-component equations to the single-fluid MHD model.  The issue of the Hall term is discussed in detail in Appendix \ref{apa}.  

Using momentum conservation $m_n n_{n0}\nu_{nj}=m_jn_0\nu_{jn}$ and $m_i=m_n$, from Eq.~(\ref{be1c}) we have the velocity  of neutrals
\be
\vec v_n\!=\!\frac{1}{\alpha} \frac{n_0}{n_{n0}} \!\left(\!\nu_{in} \vec v_i + \frac{m_e \nu_{en} }{m_n} \vec v_e\!\right),\label{vn}
\ee
\[
 \quad \alpha=\nu_{en}\frac{m_e n_0}{m_n n_{n0}} + \nu_{in} \frac{n_0}{n_{n0}} - i \omega.
\]
This is used in remaining derivations to eliminate the neutral speed.   The electron and ion momentum equations become
\[
\left[\!-i \omega m_e + m_e (\nu_{ei}+ \nu_{en}) - \frac{n_0 m_e^2 \nu_{en}^2 }{n_{n0} \alpha m_i} \! \right]\vec v_e\!=\! -e\vec E - e \vec v_e\times \vec B_0
\]
\be
+ \left(\nu_{ei} + \frac{\nu_{en} \nu_{in} n_0}{\alpha n_{n0}}\right) m_e \vec v_i, \label{ve}
\ee
\[
\left(-i \omega +   \nu_{in} + \frac{m_e \nu_{ei}}{m_i}  - \frac{n_0 \nu_{in}^2 }{n_{n0} \alpha }  \right)\vec v_i=\frac{e}{m_i}\vec E + \frac{e}{m_i} \vec v_i\times \vec B_0
\]
\be
+ \left(\nu_{ei} + \frac{\nu_{en} \nu_{in} n_0}{\alpha n_{n0}}\right) \frac{m_e}{m_i} \vec v_e. \label{vi}
\ee
 Hence, we now have a closed set of equations (\ref{bwe}, \ref{ve}, \ref{vi}) for $ \vec v_e, \vec v_i, \vec E$. This implies 5 scalar equations for $v_{ex}, v_{ey}, v_{ix}, v_{iy},  E_x$ which can be written  as:
\be
\!\!\left(\!
  \begin{array}{ccccc}
    a_1 & e B_0 & -a_2 & 0 & e \\
    -eB_0 & a_1 & 0 & -a_2 & 0 \\
    -b_2 & 0 & b_1 & -\Omega_i & -\frac{e}{m_i} \\
    0 & -b_2 & \Omega_i & b_1 & 0 \\
    ie n_0\mu_0\omega & 0 & -ie n_0\mu_0\omega & 0 & k^2 \\
  \end{array}
\!\!\!\right)
\!\!\!\left(\!
  \begin{array}{c}
    v_{ex}\\
    v_{ey} \\
    v_{ix}\\
    v_{iy} \\
    E_x \\
  \end{array}
\!\right)\!\!=\!0,
\label{s}
\ee
%Here
\[
a_1=- i\omega m_e + m_e\!\left(\nu_{ei}+ \nu_{en}\right)-\frac{n_0\nu_{en}^2 m_e^2}{\alpha n_{n0} m_i},
\]
\[
b_1=- i \omega + \nu_{in}+ \nu_{ei}\frac{m_e}{m_i} - \frac{n_0 \nu_{in}^2}{\alpha n_{n0}},
\]
\[
a_2=m_e\!\left(\!\nu_{ei} + \frac{n_0\nu_{en} \nu_{in}}{\alpha n_{n0}}\!\right)\!,\!
\quad
b_2=\frac{m_e}{m_i}\!\left(\!\nu_{ei} + \frac{n_0 \nu_{in} \nu_{en}}{\alpha n_{n0}}\!\right)\!.
\]
%The neutral speed has also two components which can be calculated from Eq.~(\ref{vn}).
 The complex frequency can be calculated by  solving  the dispersion equation
\be
\Delta(\omega, k)=0, \label{de}
\ee
 where $\Delta$ is determinant of the $5\times 5$ square matrix in  (\ref{s}) whose meaning is obvious, and $k\equiv k_z$. The dispersion equation (\ref{de}) without approximations is enormously lengthy  (see Appendix B where a simplified version of it is given) and it is of the shape $a_6\omega^6 + i a_5 \omega^5 + a_4 \omega^4 + i a_3\omega^3 + a_2\omega^2 + i a_1\omega + a_0=0$. It describes  the previous two modes [AW and forced neutral (FN) collisional mode] together with a high frequency mode due to  electron inertia, which is of no importance for the present study. We  shall deal with Eq.~(\ref{de})  numerically.

   If collisions are completely neglected the dispersion equation which follows from  complete Eqs.~(\ref{bwe}-\ref{be1b}) (after neglecting terms of the order $m_e/m_i$ with respect to unity) reads:
   \be
   \omega^4 \left(k^2 + \frac{1}{\lambda_e^2}\right) - \omega^2 \Omega_e^2 \left(k^2 + \frac{1}{\lambda_i^2}\right) + \Omega_e^2\Omega_i^2k^2=0.
   \label{comp}
   \ee
   In the regime $\omega^2\ll \Omega_i^2\ll \Omega_e^2$, $\Omega_e=eB_0/m_e$,  the first term may be neglected and this  yields the collision-less
  Alfv\'en wave spectrum  which can be written in two equivalent forms
  \be
  \omega^2\!=\!k^2 c_a^2\left(\!1-\frac{\omega^2}{\Omega_i^2}\!\right)\simeq k^2 c_a^2, \! \quad\! \mbox{or} \! \quad \!\omega^2\!=\!\frac{k^2 c_a^2}{1\!+\! k^2\lambda_i^2}\!\simeq \!k^2 c_a^2.\label{aw}
  \ee
 In the regime   $\omega^2\gg \Omega_i^2$, the last term in (\ref{comp}) may be omitted and the resulting approximate expression gives the high frequency mode
 \be
 \omega^2\simeq \Omega_e\Omega_i \frac{1+ k^2 \lambda_i^2}{1+ k^2 \lambda_e^2}. \label{hfm}
 \ee
 Frequency of this mode may exceed the electron plasma frequency and in this limit it would be appropriate to include electron density perturbations (and displacement current too) and to deal with the Langmuir and electromagnetic light modes as well. As  may be seen further in the text, derivations are extremely lengthy already and we shall keep in mind the required frequency range for this third mode which will be present in derivations, so that we avoid additional modes.

   It may also be seen that    neglecting collisions and remaining in the AW frequency range, the $x$-component of the current in Eq.~(\ref{bwe}), i.e., the polarization current, yields the speed difference
 \be
 v_{ix}-v_{ex}=\frac{i\omega}{\Omega_i} \frac{E_x}{B_0} \left(1+ \frac{m_e}{m_i}\right).\label{epol}
 \ee
 So the electron inertia term from the left-hand side of Eq.~(\ref{be1b}) is indeed negligible (yet it will be kept in the remaining calculations).

\subsubsection{\label{ns}Numerical solutions of Eqs.~(\ref{cc3},\,\ref{ni},\,\ref{de})}

\begin{figure}%[!htb]
   \centering
  \includegraphics[height=6cm,bb=17 14 266 218,clip=]{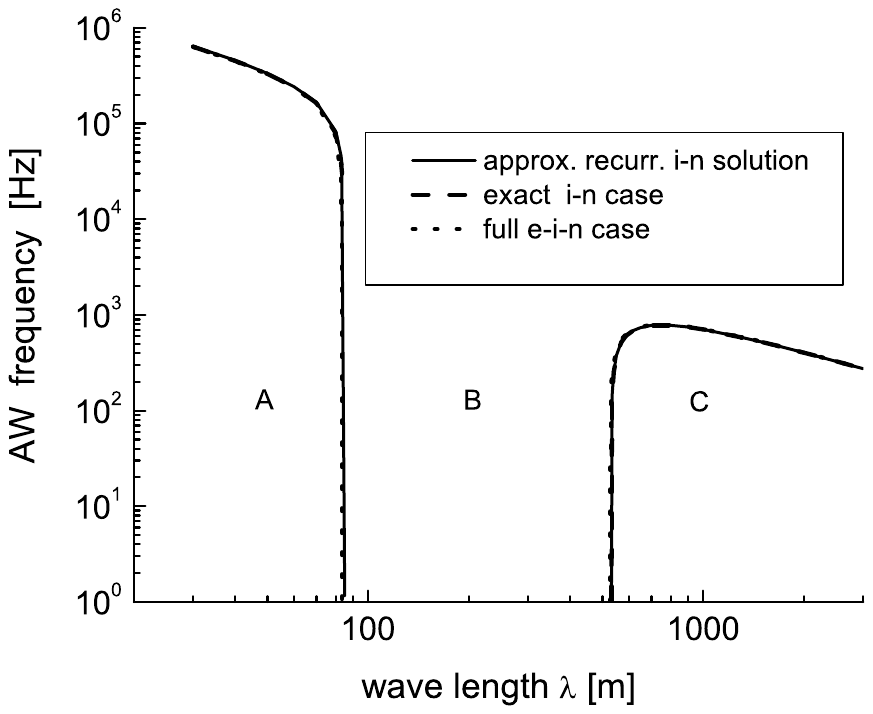}
      \caption{ \label{4-omega}Real part of frequency $\omega=\omega_r- i \gamma$ for the strongly damped Alfv\'{e}n wave as solution of
      Eqs.~(\ref{cc3},\,\ref{ni},\,\ref{de}). The result is practically the same as  Kulsrud-Pearce solution
      presented in Fig.~\ref{kp-new}. }
       \end{figure}

In what follows we shall first show that  all three equations  (\ref{cc3},\,\ref{ni},\,\ref{de})
have practically  the same solutions which for certain parameters further coincide with the Kulsrud-Pearce result shown in Fig.~\ref{kp-new}.
These three equations are solved for the same parameters used in Fig.~\ref{kp-new}, where now electron collisions are
included through Eq.~(\ref{de}) with collision frequencies $\nu_{en}\simeq 6\cdot 10^6$ Hz and $\nu_{ei}\simeq 9\cdot 10^6$
Hz [c.f., Table~\ref{teh} and also \cite{vaa}]. The result for frequencies is given in Fig.~\ref{4-omega}.
It shows that, contrary to claims in  \cite{zak}, the AW cut-off and classical results from \cite{kp}  can also
 be obtained within the fully multi-component theory.

 The differences between solutions of Eqs.~(\ref{cc3},\,\ref{ni},\,\ref{de}) are very small and not
 visible in logarithmic scales used here, which also confirms that electron contribution to the damping
 and AW mode behavior is completely negligible as correctly assumed in the approximate equations
  (\ref{cc3},\,\ref{ni}). The damping (which is not presented here) is very similar to the Kulsrud-Pearce solution given in Fig.~\ref{kp-new}.

The gap B, between the two propagation windows A and C in Fig.~\ref{4-omega}, can be controlled by several
parameters. For example, taking $B_0=0.1$  T the short wavelength cut-off $\lambda_{c1}$ is shifted towards
 $\lambda\simeq 168 $ m, and the mode re-appears again at the long wavelength cut-off $\lambda_{c2}\simeq 1050 $ m.
 However, in some cases, one of  the propagation windows  may vanish completely. However,  this behavior cannot be obtained using the Kulsrud-Pearce
 equation (\ref{kp3}), see more in Sec.~\ref{dkp}.

 \begin{figure}%[!htb]
   \centering
  \includegraphics[height=6cm,bb=15 14 274 218,clip=]{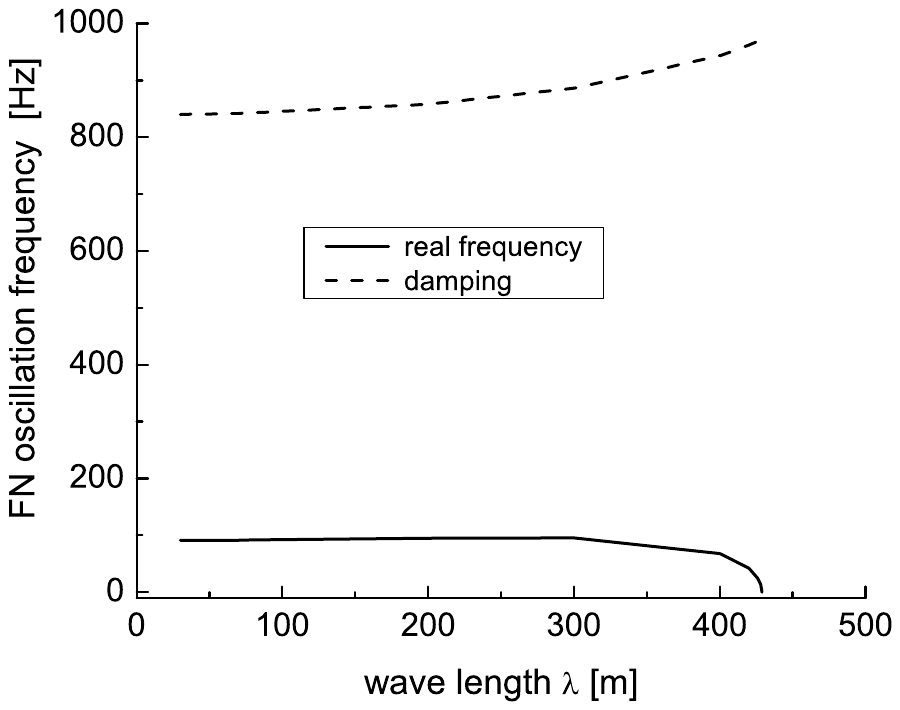}
      \caption{ \label{fn}Frequency $\omega=\omega_r- i \gamma$ of forced neutral oscillations as additional
      solution of Eqs.~(\ref{cc3},\,\ref{ni},\,\ref{de}) due to neutral friction with protons. }
       \end{figure}

In addition, the  peculiar (and highly damped) forced neutral (FN) oscillations are  presented in Fig.~\ref{fn} for the
 same parameters as in Figs.~\ref{kp-new},\,\ref{4-omega}.
 Observe  that it formally continues even for the wavelengths (up to 430 m) for which the frequency of its source (the Alfv\'{e}n wave)
  has  vanished (at around 80 m).  To explain this, note that in  the neutral equation (\ref{be1c}) there is
  no explicit dependence on the wave length; such a dependence  enters only through the ion speed which on the other
   hand is assumed  spatially varying as $\exp(i k z)$. So dependence of FN mode  on $\lambda$ is only through its
    source, and the  continuation of FN mode  above the AW critical wave length $\lambda_c$  is because  it is the
    real part of the  frequency of the AW which vanishes,  not its imaginary part of its assumed wavelength.
    So what is left of the AW is spatial variation (defined by $k$) which decreases
    exponentially in time.  The remaining imaginary part of the  AW frequency
   enters the expression for polarization drift (\ref{vi2}) and this  provides the necessary two-dimensionality in ion motion discussed
    in Sec.~\ref{arb}, so that the FN mode remains for some time even after the AW has vanished. Eventually, the FN mode vanishes as well for larger wavelengths,
    and this is due to the fact that  the larger assumed wavelength means
    more collisions within one spatial oscillation of plasma, so neutrals become better coupled to plasma oscillation
     and  have less freedom to move independently following  their own forced oscillatory mode.
 In region C it does not re-appear because of the same reason: we now have propagating AW wave and this is so only because plasma-neutrals mixture  is
 perfectly well coupled,  wave period for AW is large enough so particles from the two fluids (ions and neutrals) have time to collide many times
 in one wave period. They  move in concert perfectly well, so that neutrals do not develop their own forced (but independent) motion.

We stress again that the FN oscillation is due to two dimensional dynamics associated with the Alfv\'{e}n wave  which makes
the order of the dispersion relation higher. But the FN mode damping is too large (see Fig.~\ref{fn}) so that the mode is not expected to be observed.

All these results are merely for the demonstration, aimed at showing that the multi-component theory in principle may
yield results very similar to classic theory, \cite{kp} at least regarding the Alfv\'{e}n wave. So, contrary to recent claims,\cite{zak} keeping the Hall term in their hybrid MHD analysis makes no difference; see more on this issue in Appendix~\ref{apa}.  However, for some parameters the KP solutions may be rather different from the full multi-component theory, and this will be demonstrated in Sec.~\ref{dkp}.

\section{Application to lower solar atmosphere}

Full dispersion equation (\ref{de}) can be applied to solar atmosphere to check the existence of Alfv\'{e}n waves.  As example, for photospheric
parameters around the temperature minimum at $h=490$ km, it is  solved  in terms of the  magnetic field magnitude $B_0$ for several
 wavelengths. The densities here are \cite{fon} $n_0=2.76 \cdot 10^{15}$ m$^{-3}$, $n_{n0}=2.9 \cdot 10^{21}$ m$^{-3}$.
  Using Fig.~\ref{f1}, for the corresponding temperature $T=4410$ K we have $\sigma_{in,mt}=376.4 $ a.u. which
  yields $\nu_{in,mt}=1.84\cdot 10^7$ Hz, and for electrons $\nu_{en}=2.2\cdot 10^8$ Hz. In order to remain in the proper frequency
  range, clearly we cannot go to arbitrarily small wave lengths because $\Omega_i/\omega$ is supposed
  to be much greater than  1.

\begin{figure}%[!htb]
   \centering
  \includegraphics[height=6cm,bb=16 16 272 237,clip=]{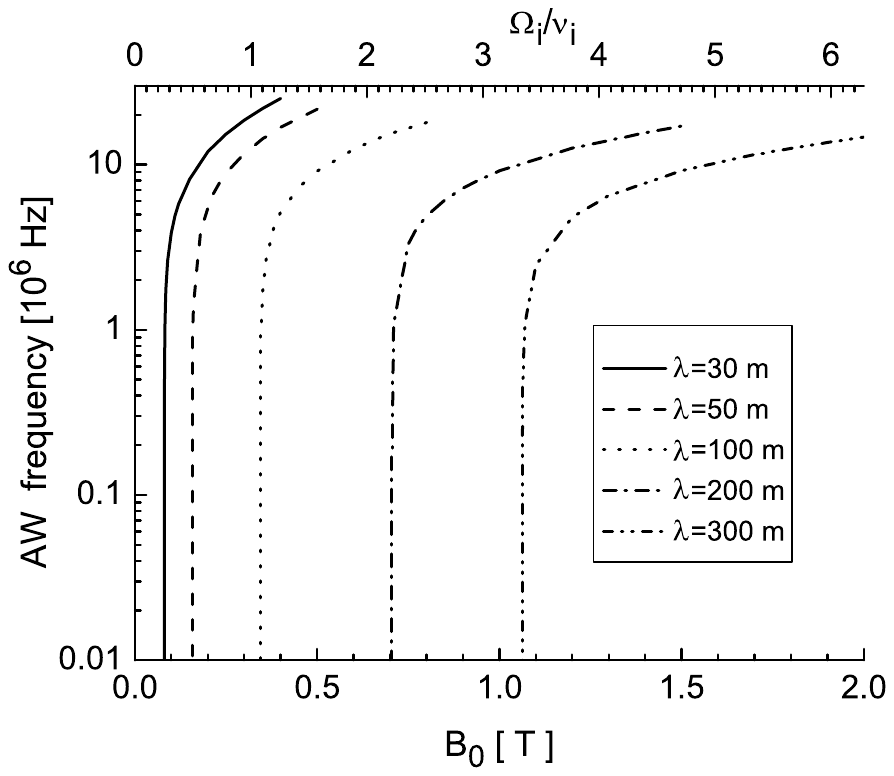}
      \caption{\label{bom200}Vanishing  of the  Alfv\'{e}n wave frequency, obtained from full dispersion equation (\ref{de}) for
       several wavelengths at altitude $h=490$ km in photosphere, in terms of magnetic field $B_0$
       (bottom axis) and ratio $\Omega_i/\nu_i$ (top axis).    }
       \end{figure}

The result for the wave lengths $\lambda = 30, 50, 100, 200, 300$ m is presented in Fig.~\ref{bom200}. For the given wave
lengths the Alfv\'{e}n wave vanishes
for the magnetic field below $B_0 \simeq 0.0804, \, 0.157, \, 0.345, \, 0.705, \, 1.063$ T, respectively. The corresponding
 wave damping is nearly constant and  for the given wave lengths it is $\gamma \simeq 1.3\cdot 10^7, 1.1\cdot 10^7 ,
 9.9\cdot 10^6, 9.45\cdot 10^6, 9.36\cdot 10^6$ Hz, respectively. Hence, not only that the wave completely vanishes below
 the given critical magnetic field magnitude, but it is also heavily damped for the magnetic field which formally allows its existence.

 Also added in the  figure as the top $x$-axis is the ratio $\Omega_i/\nu_i$ where $\nu_i\equiv \nu_{i, sc}$ is calculated using
 the total collision cross section for elastic scattering from Fig.~\ref{f1} (line 1); for the given temperature ($\simeq 0.38 $ eV)
  the cross section is $\sigma_1=622.4 \, a.u.=1.74\cdot 10^{-18}$ m$^{-2}$  so that collision frequency for total scattering
  is $\nu_{i, sc}=3.05\cdot 10^7$ Hz. It is seen that the wave vanishes for unmagnetized ions.

In application to strong magnetic structures with the starting magnetic field $B_0(0)=0.1$ T, and assuming that the
magnetic field decreases with the altitude as $B_0(x)=\exp[-x/(2 h)]$ where $h=125\cdot 10^3$ m, at the altitude
 $x=490\cdot 10^3$ m its value becomes $0.014$ T, and this is well below the required critical values given above.
  So none of the Alfv\'{e}n wave wavelengths discussed here  can be expected to appear at all, and this holds even
  if the magnetic field is kept almost constant with the altitude.

Going to  shorter wavelengths does not make much sense  because the  wave frequency becomes close
 to the gyro-frequency and the theoretical model becomes violated.
  We have checked this for $\lambda=15$ m (and for the same other parameters at $h=490$ km as above),
   and a strongly damped Alfv\'{e}n wave is obtained for the magnetic field in the range $0.04 - 0.1$ T.
   For example at $B_0=0.04$ T the AW frequency  is $\omega_r- i \gamma = 2.5\cdot 10^6- i 1.6\cdot 10^7$
    Hz while $\Omega_i=3.8\cdot 10^6$ Hz, so the theory is hardly  valid and results are unreliable,
    though even here  the wave vanishes  for  $B_0 \leq  0.028$ T.   However, in view of such a great
    damping, the Alfv\'{e}n waves in this very short wave length range are unlikely.

  \begin{figure}%[!htb]
   \centering
  \includegraphics[height=6cm,bb=16 16 270 237,clip=]{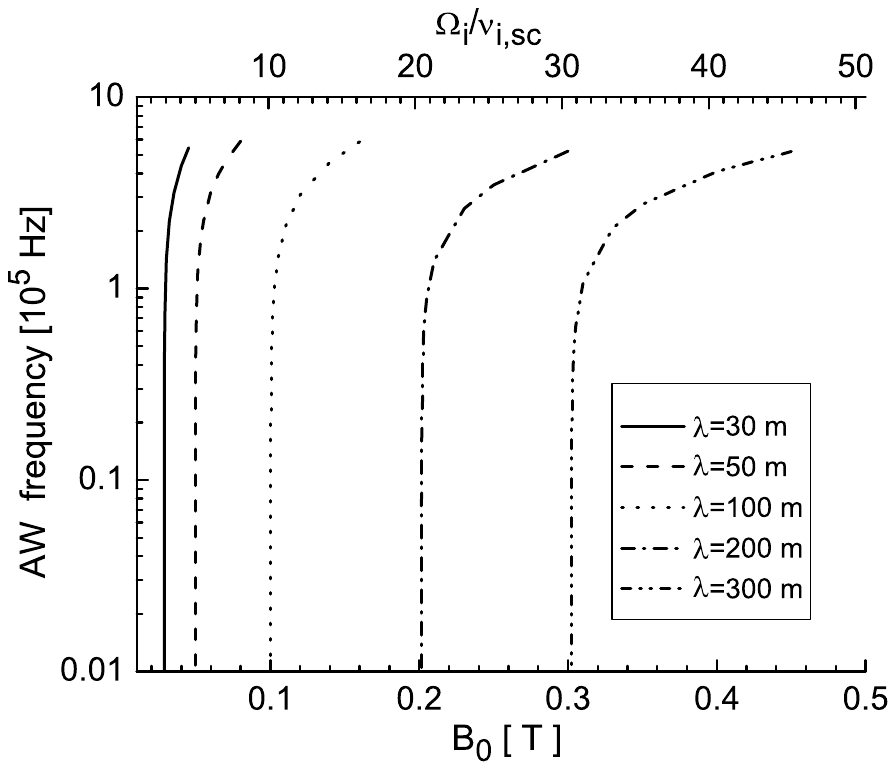}
      \caption{\label{b805}Vanishing  of the  Alfv\'{e}n wave frequency, obtained from full dispersion equation (\ref{de}) for
       several wave lengths at altitude $h=805$ km.    }
       \end{figure}

The AW  propagation for the same wave lengths is checked also at higher  altitudes, and the result  for   $h=805$ km  is
given in Fig.~\ref{b805}. The parameters are  $T=5490$ K, $n_0=8.54 \cdot 10^{16}$ m$^{-3}$, $n_{n0}=1.48 \cdot 10^{20}$ m$^{-3}$,
 and $\nu_{en}=1.3\cdot 10^7$ Hz.  Here, the wave length
$\lambda=30$ m vanishes if the  magnetic field is below $B_0\simeq 0.0285$ T, and in this case we still have magnetized protons
 because $\Omega_i/\nu_i\simeq 3$. Compare this with the wave length   $\lambda=300$ m for which the wave will not  appear if
  the the magnetic field is below  $B_0\simeq 0.3$ T, for which protons are in fact still strongly magnetized $\Omega_i/\nu_i\simeq 30$.
Here again we used the line 1 from Fig~\ref{f1} which yields $\sigma_1=657.24 \, a.u.=1.84\cdot 10^{-18}$ m$^2$ and the collision
frequency for total scattering is $\nu_{i, sc}=1.83\cdot 10^6$ Hz.
The wave damping for all given wave lengths is around $\gamma\simeq 4.7\cdot 10^5$ Hz.
Applying this again to  the strong flux tubes with the same exponential decrease as above yields the magnetic field at this
altitude around $0.004$ T only. Therefore none of the wave lengths is expected to appear.

\begin{figure}%[!htb]
   \centering
  \includegraphics[height=6cm,bb=16 17 270 237,clip=]{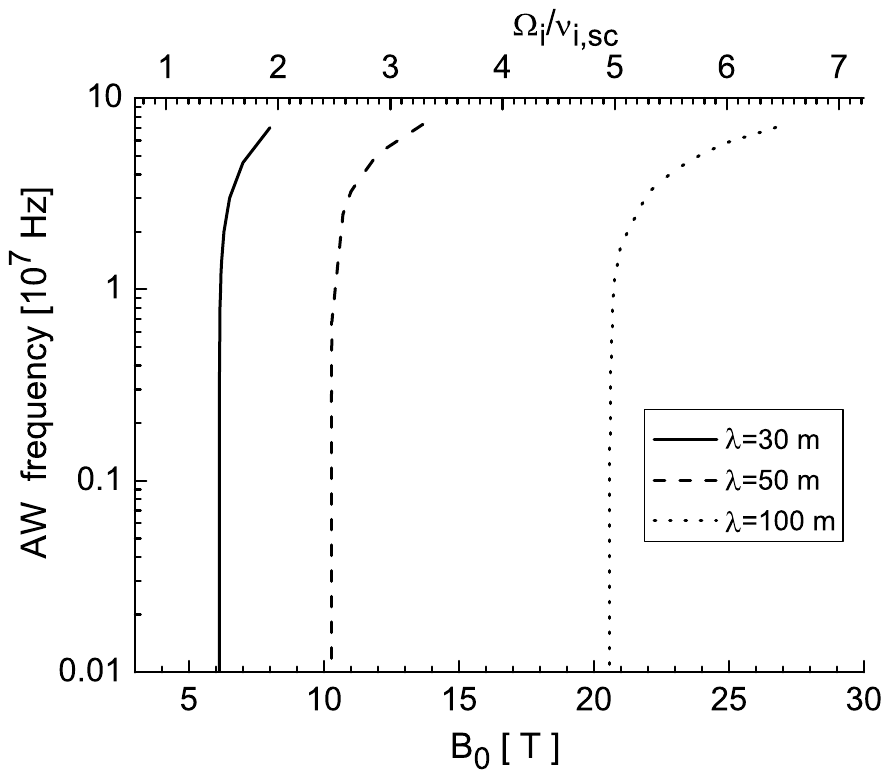}
      \caption{ \label{h200}Vanishing  of the  Alfv\'{e}n wave frequency, obtained from full dispersion equation (\ref{de}) at altitude $h=200$ km.   }
       \end{figure}

The procedure can be repeated for  the layers  below the temperature minimum. The number density of neutrals in this area
 is increased and  collision frequencies for protons \cite{vaa} go over $10^9$ Hz. Therefore in order to produce any (strongly damped)
  Alfv\'{e}n wave we need tens of T  magnetic field. For example, taking the wave length $\lambda=300$ m it turns out that the required
   magnetic field at the altitude $h=200$ km is $B_0\geq 62$ T!
The result for wave lengths $\lambda=30, \, 50, \, 100$ m is presented in Fig.~\ref{h200}. The parameters are \cite{fon}: $T=4990$ K,
$n_0=1.1 \cdot 10^{17}$ m$^{-3}$, $n_{n0}=3.47 \cdot 10^{22}$ m$^{-3}$, $\nu_{en}=2.9\cdot 10^9$ Hz, and from Fig.~\ref{f1}
this yields $\nu_{i, sc}=4\cdot 10^8$ Hz, $\nu_{in, mt}=1.7\cdot 10^8$ Hz. The AW damping for all three  wave lengths is
 around $8.5\cdot 10^7$ Hz.
It is seen that the required magnetic field for the three wave lengths are $B_0\geq 6.12, \, 10.27, \, 20.59$ T, respectively, which
 clearly shows that the  Alfv\'{e}n wave in such an environment is impossible.

Note that the given solutions  for AW are accompanied by  high frequency waves (\ref{hfm}) as well, but being unimportant for  the AW behavior
those are not presented here. As for the FN oscillations  see Sec.~\ref{dkp}.

In view of these graphs it is very unlikely that the Alfv\'{e}n wave (which is usually assumed as massively
 produced by the convective motion in the photosphere and propagating towards the corona)  can be used as a tool
 in explaining the coronal heating. From Figs.~\ref{bom200}-\ref{h200} it is seen that the wave lengths
 of several tens of meters and longer cannot possibly be excited.

We stress again that the results obtained here follow from the collisional theory summarized in Fig.~\ref{f1} which
provides the most accurate collisional cross sections, where both charge exchange and elastic scattering are consistently taken into account.

\subsection{\label{dkp}Differences in comparison with the Kulsrud-Pearce solution}

Earlier, in Figs. \ref{kp-new}, \ref{4-omega} we have demonstrated a perfect agreement (for the Alfv\'{e}n wave behavior)  between our full multi-component
model and the Kulsrud-Pearce model. However, this is not generally so, and it turns out to be dependent on parameters. The differences
in some cases are in fact profound, and with serious implications. This is shown below.

The Kulsrud-Pearce equation (\ref{kp3}) is solved for the same parameters as in Fig.~\ref{h200} and we  compare the
cases with $\lambda=100$ m.  In the range of magnetic field from Fig.~\ref{h200}, the KP solution   has very similar
behavior and vanishes for $B_c<20.43$ T, so it has  a  threshold similar to  our solutions given above. However, when the magnetic
field is further reduced, the KP solution re-appears
again at around $B_0\simeq 0.145$ T, while our AW solution does not exist below the value presented in Fig.~\ref{h200}.
Note however that our FN oscillations formally exist below this critical magnetic field value.
The FN frequency at $B_c$ is around 40 Hz and its  damping around 540 Hz.

 \begin{figure}%[!htb]
   \centering
  \includegraphics[height=6cm,bb=16 17 270 218,clip=]{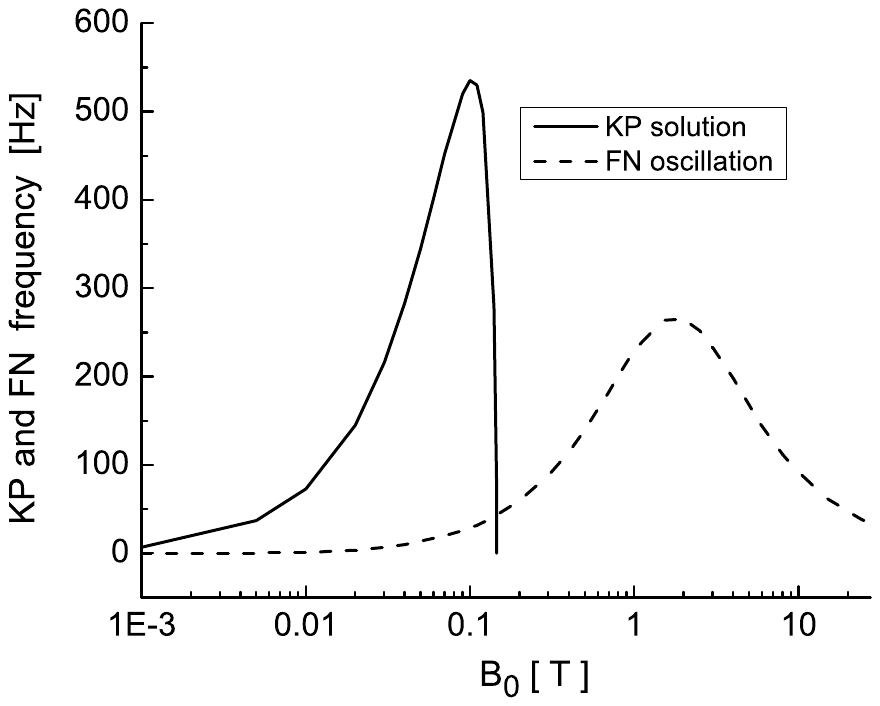}
      \caption{  \label{kpc}Second propagation window from Kulsrud-Pearce model which allows AW in the photosphere, for $\lambda=100 $ m and for parameters
      from Fig.~\ref{h200}, and our corresponding  FN mode in the same range.   }
       \end{figure}

This re-appearance of the KP solution is presented in Fig.~\ref{kpc} where we give the  real part of KP frequency
and of  our FN oscillations. The corresponding damping of the KP solution (the graph not presented here) changes from $\gamma_{kp}/\omega_{kp}\simeq 14$
(at $B_0\simeq 0.145$ T) to $\gamma_{kp}/\omega_{kp}\simeq 0.007$  at $B_0=0.001$ T. So the KP solution behaves completely
 differently as compared to our AW solution which does not exist in this magnetic field range. The  KP line shape in this range is
 a bit similar to our FN solution. In view of shortcomings of the KP equation discussed in Sec.~\ref{kp}, this KP propagation window  it is  not physical.

The similarity of the KP and our FN mode solutions is much more striking in  the following example. We take  the altitude $h=805$ km
as in Fig.~\ref{b805} and solve the KP equation (\ref{kp3}) together with our dispersion equation (\ref{de}) in terms of wave length for a fixed value
 $B_0=0.025$ T.   In the short wave length limit, both solutions are practically the same and the wave vanishes at around $\lambda=25 $ m, and this is seen
  in  Fig.~\ref{kpc2}. Note that in this wavelength range  our dispersion equation yields also the FN mode with nearly constant
  frequency and damping (see the mode presented in Fig.~\ref{kpc3} for the whole short and long wavelength range).
   Our  AW solution never re-appears again for longer wave length.
  However, the KP solution re-appears as shown in  Fig.~\ref{kpc3}. In
this long wave length range it clearly coincides with our FN mode, therefore it is not real physical  solution for the  Alfv\'{e}n wave.
It re-appears only due to earlier explained deficiencies of the MHD model.

On the other hand, the behavior of our FN mode in this wavelength range could be explained as follows. Its damping increases up to some wavelength
because of the increased amount of collisions within a FN  wave period. The increased real part of FN  frequency (up to around 450 m) is because it is driven by
two-dimensional ion motion. Yet this ion motion itself, in this range, is sustained only by the imaginary part in AW frequency which increases
with wavelength up to some critical value. After this value is achieved, the coupling between two fluids becomes more effective,
consequently the AW damping rate decreases and so does the ion two-dimensionality. This in turn affects FN frequency which therefore reduces.

\subsubsection{\label{dif}Explanation of differences}

The reason why our AW solution does not re-appear in the long wavelength range like in the case of Fig.~\ref{4-omega}
is clearly related to the magnetization and ionization ratio.
In the present case $\Omega_i/ \nu_i\simeq 2.5$ while in Fig.~\ref{4-omega} this ratio is around 10.
So now we have very weakly magnetized ions. The  AW vanishes   because  for
these larger wavelength a greater volume of neutrals must be set into motion by
 colliding ions, which themselves move due to EM perturbations. Hence,  having ions so badly magnetized their dragging is
  not so effective, neutrals thus represent a heavy obstacle and the AW vanishes.  For short wavelength there is less
   amount of collisions with neutrals within one AW wavelength and the AW propagates.

 \begin{figure}%[!htb]
   \centering
  \includegraphics[height=6cm,bb=16 14 264 218,clip=]{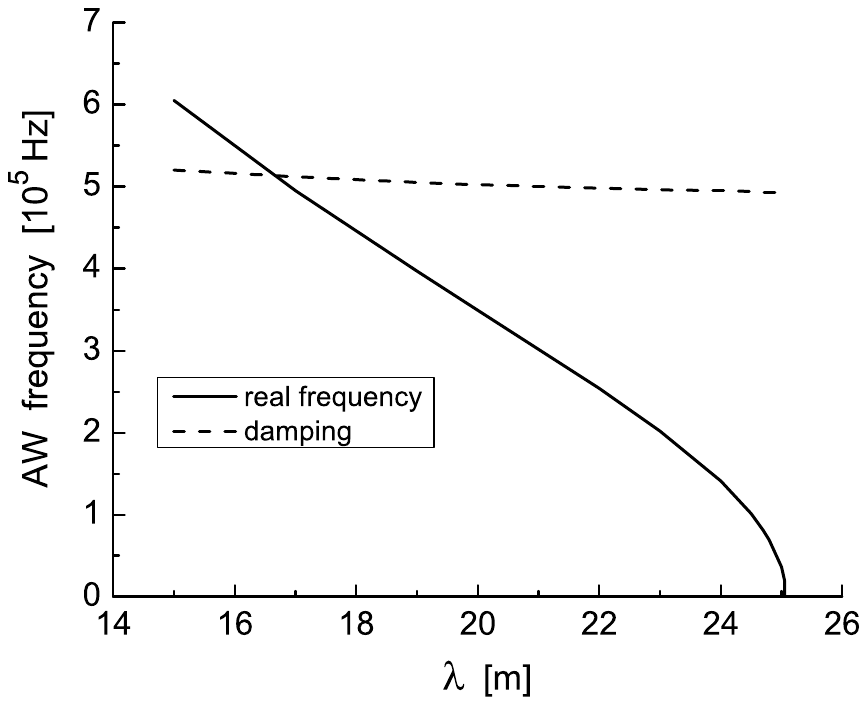}
      \caption{ \label{kpc2}Alfv\'{e}n wave frequency $\omega=\omega_r - i \gamma$
       in short wavelength range from our full multi-component model (\ref{de}) and Kulsrud-Pearce model (\ref{kp3}) at $h=805$ km and for $B_0=0.025$ T.  }
       \end{figure}

 \begin{figure}%[!htb]
   \centering
  \includegraphics[height=6cm,bb=16 14 270 214,clip=]{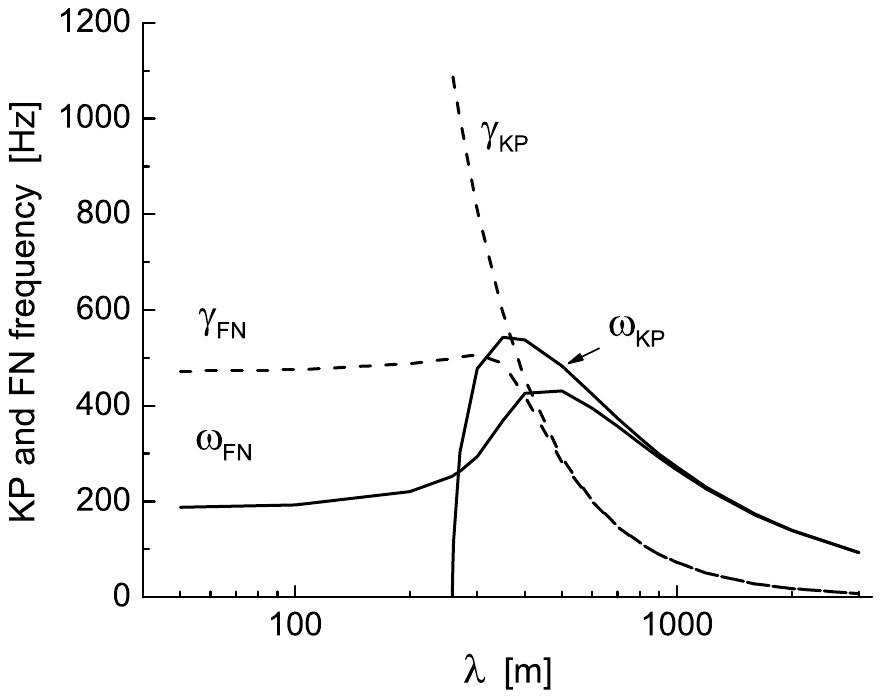}
      \caption{\label{kpc3}Complete FN mode solution $\omega=\omega_{\sss{ FN}} - i \gamma_{\sss{ FN}}$ corresponding to AW from Fig.~\ref{kpc2},
      and the Kulsrud-Pearce  long wavelength range.  }
       \end{figure}

Some more details  can be revealed  through equations in the following manner. Kulsrud-Pearce Eq.~(\ref{kp3}) can be written as:
\be
{\cal K}\equiv \left(\omega_1^2 - a^2\right) (\omega_1+ i \epsilon) + i \omega_1^2=0, \label{dd1}
\ee
\[
\omega_1\equiv\frac{\omega}{\nu_{in}}, \quad  a^2=\frac{k^2 c_a^2}{\nu_{in}^2}.
\]
%We know [c.f. Fig.~\ref{kp-new}] that in the long wavelength range Eq.~(\ref{dd1}) has positive and negative  AW as solutions, hence it   can formally be %written as
%\be
%(\omega_1+i \beta) (\omega_1-\omega_a)(\omega_1+ \omega_a)=0,\label{dd2}
%\ee
%where $\omega_a$ is the AW frequency and $\beta$  is some value which describes the existing purely imaginary solution.
%
%
We have already shown that our all three equations (\ref{cc3}, \ref{ni}, \ref{de}) yield the same solutions, so it is enough to
discuss the simplest one, Eq.~(\ref{cc3}), which can be written as
\be
\left(\omega_1^2 - a^2\right)\! (\omega_1+ i \epsilon)^2\! + i \omega_1^2(\omega_1+ i s)\!=0, \!\quad s=\epsilon + k^2 \lambda_i^2. \label{dd3}
\ee
For parameters from Figs. \ref{kp-new},~\ref{4-omega} we have $\epsilon=0.0016$, while  $k^2 \lambda_i^2=0.00007$ in the beginning of the
second propagation region C. So $s\simeq \epsilon$, and Eq.~(\ref{dd3}) for these parameters becomes of the shape:
\[
{\cal K}\cdot (\omega_1+ i \epsilon)=0.
\]
Hence, our equation clearly has the same or similar AW as solution (with the extra terms which in the end yields the FN mode). This may be seen more clearly
for large wavelengths after neglecting $a^2$ and using the fact that  $\epsilon\ll 1$ when Eq.~(\ref{dd3}) reduces to
\be
\omega_1^2 \left(\omega_1^2 + i \omega_1 -1\right)=0. \label{dd4}
\ee
This equation clearly has two real nontrivial solutions that can only be associated with AW because FN mode is shown numerically to be
absent in this wavelength range, and we thus have an agreement with the Kulsrud-Pearce case.

However, for parameters from Figs.~\ref{kpc2}, \ref{kpc3} we have  $\epsilon=0.00058$, while  $k^2 \lambda_i^2=0.0006$,  $k^2 \lambda_i^2= 0.00027$ for wavelengths
$\lambda=200, 300$ m where KP mode re-appears, so $k^2 \lambda_i^2$ cannot be omitted  and Eq.~(\ref{dd3}) is  written as
 \be
 (\omega_1+ i \epsilon)\cdot {\cal K} - \omega_1^2 k^2 \lambda_i^2=0.\label{dd5}
 \ee
We see now that it is the ion inertial length term $k^2\lambda_i^2$ which makes the difference. It naturally appears in multi-component
  theory  while it is  absent in KP MHD description [see Eq.~(\ref{kp3})].   If it is omitted in our Eqs.~(\ref{cc3}, \ref{ni}, \ref{de}) we obtain
  AW behavior similar to  KP model, with two propagation windows.
Parameters used for Figs.~\ref{kpc2},~\ref{kpc3} show that it may be greater than  the other small term $\epsilon=n_0/n_{n0}$ in Eq.~(\ref{kp3})
and consequently it cannot always be omitted. Its complete absence in MHD theory has serious consequences: this theory fails to predict the forced neutral mode,
and as a result it  allows AW in the environment where it cannot exist.  We stress that the  parameter $k^2\lambda_i^2$ which changes the physics is not introduced on purpose; it just  follows from the multicomponent theory in its simplest shape. It is an intrinsic feature of this theory which reflects essential differences between the usual MHD and our derivations.

We can now check the validity of the formula (\ref{fnn}) describing the FN mode.
For parameters from Figs.~\ref{kpc2},~\ref{kpc3}, and taking  wavelength $\lambda=600$ m,  numerical solution of  dispersion equation (\ref{de})
is   $\omega_{{\sss FN}}=397$ Hz while
the formula (\ref{fnn}) yields 444 Hz. For $\lambda=2000$ m the two values are, respectively, 139 Hz and  140 Hz. So the agreement appears
to be perfect at long wavelengths. On the other hand, the KP results for the two wavelengths are, respectively, 424 Hz and 139.7 Hz. The formula (\ref{fnn})
which describes the neutral mode is obtained after explicitly neglecting terms which yield the Alfv\'{e}n wave, so its perfect agreement with the KP results is
a direct proof that the second propagation window in Kulsrud-Pearce model is not actual physical Alfv\'{e}n wave. It is in fact associated with the neutral mode
as the multi-component theory predicts, yet this cannot be seen from the MHD theory.

\section{Conclusions}
Full three component analysis given in this work yields results that are partly in agreement with the classical MHD theory\ cite{kp}
 and with more recent ones \cite{pud, bb} based on the MHD theory.  The classic analysis  \cite{kp} gives two different
  regimes for propagation of Alfv\'{e}n waves, first  where the wave damping is proportional to the collision frequency, and second  which
  implies an inverse proportionality. Our multi-component analysis in principle confirms such a behavior, contrary to recent claims in 
   \cite{zak} that the AW cut-off must vanish if Hall term is included.   We have also identified some forced oscillations of the
    neutral fluid caused by friction with plasma species, which  cannot be obtained from the MHD  analysis.

    However, the agreement of our analysis with  \cite{kp} depends on particular plasma
    parameters, and this is demonstrated in Figs.~\ref{kp-new}, \ref{4-omega} from one side (where the agreement is perfect),
    and in Figs.~\ref{h200}, \ref{kpc}
    (or in Figs.~\ref{kpc2}, \ref{kpc3}) from the other, where some essential differences appear, and those are caused by some intrinsic deficiencies
    of the  MHD model.
From these  figures it may be concluded that the second propagation window from the Kulsrud-Pearce model (which is correctly described
for parameters in Fig.~\ref{kp-new}), for some other parameters may become un-physical (as it is the case with Fig.~\ref{kpc} and with  Fig.~\ref{kpc3}).
The origin of differences is identified:  MHD analysis in weakly ionized environment involves one small parameter $\epsilon=n_0/n_{n0}$ while
in the same time it misses other small parameter $k^2\lambda_i^2$, which may be of the same order and which is naturally included only through multi-component
theory. As an old subject,  the Alfv\'{e}n wave has already been studied experimentally in numerous works in the past, see for example \cite{gig}, \cite{vin}, and in particular \cite{wat} and \cite{gek} dealing with experimental partially ionized plasma; in the present work we have delivered a lot of results that should be kept in mind in eventual future experiments.

From our analysis it follows also that speaking about the Alfv\'{e}n wave in an environment where ions are un-magnetized is not justified.
This is shown partly analytically in an approximate derivations, and numerically by solving dispersion equation (\ref{de}) without any approximation.
Therefore, the Alfv\'{e}n wave cannot be excited in an environment like the solar photosphere and this is demonstrated by using specific photospheric parameters
and the most accurate collision cross sections that exist. Note that this contradicts the Kulsrud-Pearce model which yields the AW in Fig.~\ref{kpc} in the range
where ions are un-magnetized. This shows that the most popular paradigm of the coronal heating by Alfv\'{e}n waves  produced in the photosphere is against physical reality: the ions in
the photosphere are unmagnetized \cite{vaa} and they cannot support the Alfv\'{e}n wave.

Regarding the work  \cite{zak} where it is claimed that the AW cut-off is not possible, at this point  it is appropriate to make
the following comments. They introduce relative perpendicular speed between ions and neutrals $\vec w_{\bot}=\vec u_{i\bot}-\vec u_{n\bot}$,
   and the common speed of the two (ions plus neutrals) fluids $\vec u_{\bot}$.
    However,  they explicitly neglect the time derivative of the relative speed  $ \partial \vec w_{\bot}/\partial t$, while such a time
    derivative for the common speed is kept.  By neglecting the time variation of the relative speed they have directly  {\em excluded}
     physical phenomena   which develop within  transition (collisional) time, and  which dictate everything what happens with the mixture
     of the two fluids (plasma plus neutrals). In other words,  the relative motion of neutrals and ions they  assume fixed in time. By
     doing this they prevent the system to evolve freely, and this partly removes effects of friction.    Such an assumption is physically
     unjustified and in view of this it is no surprise that they
     do not obtain any cut-off.  It is also very likely that the FN mode identified in our work is simply overlooked  in their work and
     interpreted as a low frequency continuation of the Alfv\'{e}n wave (hence the absence of cut-off in their work). Namely,  they
     derive dispersion equation with  a free term $\delta_i \nu \xi_i$ (in their own notation), which seems to be equivalent to the free term in
      our derivation where it yields the FN oscillations.

{\bf Acknowledgments:} JV is enormously grateful to P. S. Krstic for help and valuable discussions related to the calculation and understanding  of collision cross sections which involve quantum-mechanical indistinguishability of colliding particles at low energies.

\appendix

\section[\label{apa}On the role of the Hall effect in MHD model ]{\label{apa}On the role  of the Hall effect in MHD model }

The Hall effect belongs completely to the MHD terminology and it can introduce some new phenomena  only within this model, and this when used instead of the
 ideal Ohm's law of course.  However, the fully multi-component plasma theory  used in the present work, being more general, contains all effects that are
within the MHD theory normally attributed to the MHD Hall effect, and this will be demonstrated here.

   The generalized MHD Ohm's law, which contains the Hall term, is obtained by combining and re-arranging the momentum equations for plasma components.
    Our three momentum equations from Sec.~\ref{b}     can be combined in various ways to obtain the  generalized Ohm's
    law. Observe that within the multi-component plasma theory such a combined equation is redundant because we operate with velocities of separate species
    and not with the current. We  may multiply our ion and electron momentum equations   by $e/m_i$ and $-e/m_e$, respectively, and then sum  the resulting two equations assuming  quasi-neutrality, which yields
\[
e n (\vec v_i-\vec v_e)=\sigma \vec E+\underline{\frac{\sigma}{\beta}\left(\frac{e^2 n}{m_i}\vec v_i+\frac{e^2 n}{m_e}\vec v_e\right)\times \vec B}
\]
\[
+
\frac{e \sigma}{\beta}\nabla\left(\frac{p_e}{m_e}- \frac{p_i}{m_i}\right)-\frac{\sigma}{\beta}\frac{\partial}{\partial t}\left[ e n (\vec v_i-\vec v_e)\right]
\]
\[
-\frac{e n\sigma}{\beta} \left[\left(\vec v_i\cdot\nabla\right)\vec v_i-\left(\vec v_e\cdot\nabla\right)\vec v_e\right]
\]
\be
+\frac{e n\sigma}{\beta} \left[\nu_{en} \left(\vec v_e-\vec v_n\right)- \nu_{in} \left(\vec v_i-\vec v_n\right)\right].
\label{apa1}
\ee
Here
\[
  \sigma=\frac{e^2n}{m_e\nu_{ei}}, \quad \beta=\frac{e^2 n}{m_i}+ \frac{e^2 n}{m_e}.
\]
The underlined part of the equation comes from the Lorentz force terms for both electrons and ions, and this part contains  the usual MHD Hall term. Indeed, after introducing the total speed of the fluid
$(m_i+ m_e)\vec V =m_i n\vec v_i+ m_e n \vec v_e$, and neglecting only the ion contribution in $\beta$ (due to mass difference), with simple transformations Eq.~(\ref{apa1}) becomes
\[
\overbrace{e n (\vec v_i-\vec v_e)}^{\vec j}= \sigma \left(\vec E + \vec V\times \vec B\right) + \frac{e}{m_e\nu_{ei}} \nabla \left(p_e-\frac{m_e}{m_i} p_i\right)
\]
\[
-
\underline{\frac{e}{m_e \nu_{ei}} \overbrace{e n (\vec v_i-\vec v_e)}^{\vec j}\times \vec B}- \frac{1}{\nu_{ei}} \frac{\partial \overbrace{e n (\vec v_i-\vec v_e)}^{\vec j}}{\partial t}
\]
\[
-\frac{e n}{\nu_{ei}} \left[\left(\vec v_i\cdot\nabla\right)\vec v_i-\left(\vec v_e\cdot\nabla\right)\vec v_e\right]
\]
\be
 +
\frac{e n}{\nu_{ei}}  \left[\nu_{en} \left(\vec v_e-\vec v_n\right)- \nu_{in} \left(\vec v_i-\vec v_n\right)\right]. \label{apa2}
\ee
Further simplifications are clearly possible but this is not necessary to do because we already see that  this equation is the generalized Ohm's law, and the underlined term in Eq.~(\ref{apa2}) is the well-known MHD Hall term. The origin of the Hall term and the $\vec V\times \vec B$ term  is in the Lorentz force. This equation contains the velocities of ions and electrons, and it must be complemented with any of the two used momentum equations
in order to have a closed set of equations. Such a new set will  again contain exactly the same physics, and it  will again yield
the same dispersion equation given in the Appendix B.

This manipulation  with momentum equations can be done differently, by adding all three momentum equations (for electrons, ions and neutrals).
 But as above, this equation must then be complemented by two of the starting three equations to close the set, and the resulting dispersion equation will again be  the same as if this manipulation of momentum equations is not used at all.

 It may be
concluded that our multi-component equations contain all physics equivalent to the Hall effect within the MHD theory.  Nevertheless,
with all this  we are  still
able to recover the classic KP results as shown in Fig.~\ref{4-omega} in Sec.~\ref{ns}.
This additionally shows that recent claims \cite{zak} of new phenomena introduced by the Hall effect, which apparently remove the classic  KP cut-off, cannot possibly be correct,  and the Hall effect does not change the classic Kulsrud-Pearce result. Reality is that  the essential new phenomena which make the difference arise from the fact that the more complete
dispersion equation derived in our work contains additional small parameter with the ion inertial length, and this is unrelated to  the Hall term.

%\appendix
\section[Dispersion equation for collisional Alfv\'{e}n wave in three-component plasmas]{Dispersion equation for collisional Alfv\'{e}n wave in three-component plasmas}

Dispersion equation (\ref{de}) in explicit form reads:
\be
a_6\omega^6 + i a_5 \omega^5 - a_4 \omega^4 - i a_3\omega^3 + a_2\omega^2 + i a_1\omega - a_0=0.\label{a1}
\ee
After neglecting only terms with  $m_e/m_i$ with respect to 1 (some additional simplifications are clearly possible), the coefficients read:
\[
a_0=\Omega_i^2\Omega_e^2 k^2 \lambda_e^2 \frac{n_0^2}{n_{n0}^2}  \left(\nu_{in} + \nu_{en} \frac{m_e}{m_i}\right)^2, \quad \lambda_e=\frac{c}{\omega_{pe}},
\]
\[
a_1= 2 k^2\lambda_e^2 \Omega_i^2 \frac{m_e^2}{m_i^2} \frac{n_0^2}{n_{n0}^2} \left(\nu_{en} + \nu_{in} \frac{m_i}{m_e}\right)\left\{ \nu_{ei} \nu_{en}
 \right.
\]
\[
  \,\,\,\,\,\left. + \Omega_e^2 \frac{m_i}{m_e} \frac{n_{n0}}{n_0} +  \frac{m_i^2}{m_e^2} \left(1+ \frac{n_{n0}}{n_0}\right) \nu_{in} (\nu_{ei}+ \nu_{en})
\right.
\]
\[
 \,\,\, \,\,\,\,\,+ \frac{m_i}{m_e}\left[\nu_{en}\nu_{in} + \nu_{ei} (\nu_{en}+ \nu_{in}) \right.
\left.\left.
+ \nu_{ei} \nu_{en} \frac{n_{n0}}{n_0}\right]\right\},
\]
\[
a_2=\Omega_i^2 \frac{n_0^2}{n_{n0}^2} \left[\!\left(\!\nu_{en}^2 \frac{m_e}{m_i}\!  +\! 2 \nu_{en} \nu_{in}\right) \left(1\!+ \! \frac{n_{n0}}{n_0}\right)
\right.
\]
\[
\left. \,\,\, \,\,\,\,+ \!\nu_{in}^2\! \left(\!1+ \!\frac{m_i}{m_e}\!\left(\!1+ \!\frac{n_{n0}}{n_0}\!\right)\!\right)\!\right]
 \!+\! \Omega_i^2 k^2 \lambda_e^2\left\{ \!\Omega_e^2\! \right.
\]
\[
\,\,\, \,\,\,\, +\!2 \nu_{ei} \nu_{en} \left(\!1\!+\! \frac{2n_0}{n_{n0}}\! \right) +\!
\left.\nu_{en}^2 \left[\!\left(\!1\!+\! \frac{m_e}{m_i} \frac{n_0}{n_{n0}}\!\right)^2\! +\!\frac{n_0^2}{n_{n0}^2}\! \right]\right.
\]
\[
\left.
\!+\! \nu_{in}^2 \left[\!\frac{m_i^2}{m_e^2}\left(\! 1\!+\! \frac{n_0}{n_{n0}}\! \right)^2\! +\! \frac{n_0^2}{n_{n0}^2}\!\right]
\right.
\]
\[
\,\,\, \,\,\,\, \left. +\!2 \nu_{en} \nu_{in} \frac{n_0^2}{n_{n0}^2}\! \left[\!\frac{m_i}{m_e} \!\left(\! 1\!+\! \frac{2n_{n0}}{n_0}\!\right)
\!+\! \frac{2n_{n0}}{n_0}\!\right]\! \!+ \!4 \nu_{ei} \nu_{in} \frac{m_i}{m_e}   \frac{n_0}{n_{n0}}\!\right\}
\]
\[
\,\,\, \,\,\,\,\,+ k^2 \lambda_e^2 \left\{ \nu_{in}^2 \left(\nu_{en}^2+  \nu_{ei}^2\right) \left(1+ \frac{n_0}{n_{n0}}\right)^2
\right.
\]
\[
\left.
\,\,\, \,\,\,\,\,+\nu_{ei}^2 \nu_{en}^2\frac{m_e^2}{m_i^2} \left[\left(1+ \frac{2m_e}{m_i}\frac{n_0^2}{n_{n0}^2} + \frac{n_0^2}{n_{n0}^2}\right) +
2 \frac{n_0}{n_{n0}} \right]
\right.
\]
\[
\left.
\,\,\, \,\,\,\,\,+ 2\nu_{ei} \nu_{in} \nu_{en}^2 \frac{m_e}{m_i} \frac{n_0}{n_{n0}}  \left( 2 + \frac{n_0}{n_{n0}}
\right)
\right.
\]
\[
\left.
 \,\,\,\,\,+ \frac{2m_e}{m_i} \left(\nu_{en}\nu_{in}\nu_{ei}^2\! +\! \nu_{ei} \nu_{in}\nu_{en}^2\right) + 2 \nu_{ei}\nu_{en}\nu_{in}^2 \left(\!1+
\frac{n_0^2}{n_{n0}^2}\! \right)
\right.
\]
\[
\left.
 \,\,\,\,\,+  2 \nu_{en}\nu_{in}\nu_{ei}^2 \frac{m_e}{m_i}\frac{n_0}{n_{n0}}  \left( 4 + \frac{n_0}{n_{n0}} \right)
+ \frac{4n_0}{n_{n0}}  \nu_{ei}\nu_{en}\nu_{in}^2\right\},
\]
\[
a_3=\Omega_i^2\left\{ \left(\nu_{en} + \nu_{in}  \frac{m_i}{m_e}\right) \left(1+  \frac{2n_0}{n_{n0}} \right)
 + 2 k^2 \lambda_e^2 \left[ \nu_{ei}  \frac{m_i}{m_e}   \right.
    \right.
\]
\[
\left.\left.
+   \nu_{en} \left(1+   \frac{m_i}{m_e}\frac{n_0}{n_{n0}} \right)+ \nu_{in} \left(\frac{n_0}{n_{n0}} +
\frac{m_i^2}{m_e^2}\left(1+  \frac{n_0}{n_{n0}}\! \right)\!\right)\!\right]\!\right\}
\]
\[
 + (\nu_{ei}+ \nu_{en})\nu_{in}^2 + \nu_{en}^2 \frac{m_e}{m_i} \left[ \nu_{in}+ \nu_{ei} \frac{m_e}{m_i}
\left(1+  \frac{n_0^2}{n_{n0}^2} \right)\right]
\]
\[
+ 2 \nu_{ei} \nu_{en}^2 \frac{n_0}{n_{n0}}  \frac{m_e^2}{m_i^2}\left(1+  \frac{n_0}{n_{n0}}\frac{m_e}{m_i}\right) +
\nu_{in} \nu_{en}^2 \frac{m_e}{m_i} \left(\!1\!+\!  \frac{2n_0}{n_{n0}}\!\right)
\]
\[
 +\! 2 \nu_{ei} \nu_{en} \nu_{in}  \frac{m_e}{m_i}\! \left(\!1+\! \frac{n_0}{n_{n0}}\!\right)^2\! +\!
(\nu_{ei}\!+\! \nu_{en}) \nu_{in}^2 \frac{n_0}{n_{n0}}\! \left(\!2\!+\! \frac{n_0}{n_{n0}}\!\right)
\]
\[
  + 2 k^2\lambda_e^2 \left\{ \frac{m_e}{m_i} (\nu_{ei}+ \nu_{en}) \nu_{ei}\nu_{en} + \nu_{in} \nu_{ei}^2  + \nu_{in} \nu_{en}^2
 +\nu_{ei} \nu_{in}^2
     \right.
 \]
\[
\left.  + \!\nu_{en} \nu_{in}^2\! +\! 2 \nu_{ei}\nu_{en}\nu_{in}\! +\!  \frac{n_0^2}{n_{n0}^2}\left[\! \nu_{en}\nu_{in}^2
+   \nu_{ei}\nu_{in}^2  + \nu_{in} \nu_{en} \frac{m_e}{m_i} \right.
\right.
\]
\[
\left.\left.+ \nu_{ei} \nu_{en}^2 \frac{m_e^2}{m_i^2}\!+ \! 2 \nu_{ei} \nu_{en} \nu_{in} \frac{m_e}{m_i}\!\right]
\!\!+\!\! \frac{2n_0}{n_{n0}}\!\! \left[\!\nu_{ei}\nu_{en}\nu_{in}\!+\!  \nu_{in}^2 (\!\nu_{ei} \! +\! \nu_{en}\!)\!\right]
\right.
\]
\[
\left. + \nu_{in} \left(\nu_{ei}^2+ \nu_{en}^2\right)\frac{n_0}{n_{n0}}+
\nu_{ei} \nu_{en} (\nu_{ei}+ \nu_{en} )\frac{m_e}{m_i}\frac{n_0}{n_{n0}}\right\},
\]
\[
a_4=\Omega_e \Omega_i\!+\! \Omega_e^2 k^2 \lambda_e^2 + \nu_{in}^2\! \left(\!1+ \!\frac{n_0}{n_{n0}}\!\right)^{\!2}\!
+2 \nu_{ei} \nu_{en} \frac{m_e}{m_i}\! \left(\!1+ \!\frac{n_0}{n_{n0}}\!\right)
\]
\[
 \,\,\,\,\,+  \nu_{en}^2 \frac{m_e}{m_i}  \left[1+ \frac{n_0}{n_{n0}} \left(1+\frac{m_e}{m_i} \frac{n_0}{n_{n0}}\right)\right]
+  2 \nu_{ei} \nu_{in}  \left(1+ \frac{n_0}{n_{n0}}\right)
\]
\[
 \,\,\,\,\,+ 2\nu_{en} \nu_{in}\left[1+ \frac{n_0}{n_{n0}} \left(1+ \frac{m_e}{m_i} \frac{n_0}{n_{n0}}\right)\right]
\]
\[
  + k^2\lambda_e^2 \!\left\{\! \nu_{ei}^2 \! +\! \nu_{in}^2\! \left(\!1+ \!\frac{ n_0}{n_{n0}}\!\right)^2 \!\!+\!
\nu_{en}^2\! \left[\!1+ \! \frac{m_e}{m_i} \frac{ n_0}{n_{n0}}\left(\!2  \right.\! \right.\right.
\]
\[
 \left.
 + \left.\!\frac{m_e}{m_i} \frac{ n_0}{n_{n0}}\!\right)
 \!+\! 2 \nu_{ei} \nu_{en}\left(\!1+ \frac{2m_e}{m_i}\frac{ n_0}{n_{n0}}\!\right)\! + \!4 \nu_{ei} \nu_{in} \left(\!1+ \frac{ n_0}{n_{n0}}\!\right) \right.
\]
\[
\left.+ 2 \nu_{en} \nu_{in}\left[ 2+ \frac{ n_0}{n_{n0}}\left(2+ \frac{m_e}{m_i} \frac{ n_0}{n_{n0}} \right)\right]     \right\},
\]
\[
a_5=\nu_{ei}\left(\!1+  2 k^2 \lambda_e^2\right) + 2\nu_{in} \left[\! 1+  \frac{ n_0}{n_{n0}} +
 k^2 \lambda_e^2 \left(1+ \frac{n_0}{n_{n0}}\!\right)\!\right]
 \]
 \[
  \,\,\, \,\,\,\,\,+ \nu_{en} \left[1+ \frac{2m_e}{m_i}\frac{n_0}{n_{n0}}  +
2 k^2 \lambda_e^2 \left(1+ \frac{m_e}{m_i}\frac{n_0}{n_{n0}}\right)\right],
\]
\[
a_6=1+  k^2\lambda_e^2.
\]
We stress that in the present work we have used full dispersion equation (\ref{de}) without approximations (with complete electron contribution), instead of Eq.~(\ref{a1}) with the approximate coefficients $a_j$ given above.


\begin{thebibliography}{}
\bibitem{kp} R. Kulsrud and W. P. Pearce,  Astrophys. J. {\bf 156}, 445 (1969).

\bibitem{pud} R. E. Pudritz,   Astrophys. J. {\bf 350}, 195 (1990).

\bibitem{kum} N. Kumar  and B. Roberts, Solar Phys. {\bf 214}, 241 (2003).
\bibitem{sol} R. Soler, M. Carbonell, J. L. Ballester, and J. Terradas,  Astrophys. J. {\bf 767}, 171 (2013).


\bibitem{ts} Y. T. Tsap, A. V. Stepanov, and Y. T. Kopylova,   Solar Phys. {\bf 270}, 205 (2011).
\bibitem{v1} J. Vranjes,  S. Poedts, B. P. Pandey, and B. De Pontieu,  Astron. Astrophys. {\bf 478}, 553 (2008).
\bibitem{zak}  T. V. Zaqarashvili, M. Carbonell, J. L.  Ballester, and M. L. Khodachenko, Astron. Astrophys. {\bf 544}, A143 (2012).

 \bibitem{kr1}  P. S. Krstic and  D. R. Schultz,  {\em Atomic and Plasma-Material Interaction Data for Fusion, Vol. 8} (IAEA, Vienna, 1998).

\bibitem{kr2}   P. S. Krstic and  D. R. Schultz, J. Phys. B:  At. Mol. Opt. Phys. {\bf 32}, 3485 (1999).

\bibitem{kr3}   P. S. Krstic and  D. R. Schultz, Phys. Rev. A {\bf 60}, 2118 (1999).

\bibitem{vaa} J. Vranjes and P. S. Krstic, Astron. Astrophys. {\bf 554}, A22 (2013).

\bibitem{glas} A. E. Glassgold,  P. S. Krstic and  D. R. Schultz,  Astrophys. J. {\bf 621}, 808 (2005).



\bibitem{schul}  D. R. Schultz, P. S. Krstic, T. G.  Lee, and J. C.  Raymond, Astrophys. J. {\bf 678}, 950 (2008).
\bibitem{reiz} Y. P. Raizer, {\em Gas Discharge Physics} (Springer-Verlag, Berlin Heidelberg, 1991). %157
\bibitem{varg} N. Vargaftik, Y. K. Vinogradov, and V. S. Yargin,   {\em Handbook of Physical
Properties of Liquids and Gasses} (Begell House, New York, 1996).
\bibitem{dal} A. Dalgarno, M.  Yan, and W. Liu,  Astrophys. J. Suppl. {\bf 125}, 237 (1999).
\bibitem{ch} F. F. Chen, {\em Introduction to Plasma Physics and Controlled Fusion} (Plenum Press, New York, 1988). %197
\bibitem{cat} P. J. Catto,  Phys. Plasmas {\bf 1}, 1936 (1994).
\bibitem{hel} P. Helander, S. I. Krasheninnikov, and P. J. Catto, Phys. Plasmas {\bf 1}, 3174 (1994).
\bibitem{bel} P. M. Bellan,  Phys. Plasmas {\bf 1}, 3523 (1994).
\bibitem{v10} J. Vranjes and S. Poedts,  Phys. Plasmas {\bf 17}, 022104 (2010).
\bibitem{gg} W. G. Roberge and  G. E.  Ciolek, MNRAS {\bf 382}, 717 (2007).
\bibitem{fon} J. M. Fontenla, E. H. Avrett, and R. Loeser,  Astrophys. J. {\bf 406}, 319 (1993).
\bibitem{bb} O. M. Blaes and  S. A.  Balbus,  Astrophys. J. {\bf 421}, 163 (1994).
\bibitem{gig} A. Gigliotti, W. Gekelman, P. Pribyl, S. Vincena, A. Karavaev, X. Shao, A. Surjalal Sharma, and D. Papadopoulos, Phys. Plasmas {\bf 16}, 092106 (2009).
\bibitem{vin} S. T. Vincena, G. J. Morales, and J. E. Maggs, Phys. Plasmas {\bf 17}, 052106 (2010).
\bibitem{wat} C. Watts and J.  Hanna, Phys. Plasmas {\bf 11}, 1358 (2004).
\bibitem{gek} W. Gekelman, S. Vincena, B. Van Compernolle, G. J. Morales, J. E. Maggs,
P. Pribyl, and T. A. Carter,  Phys. Plasmas {\bf 18}, 055501 (2011).
\end{thebibliography}
\end{document}